\documentclass[12pt]{article}

\usepackage{import}
\usepackage{titling}
\usepackage{xcolor}
\usepackage{authblk}
\usepackage{graphicx}
\usepackage{amsmath}
\usepackage{units}

\usepackage{bbm}
\usepackage{bm}
\usepackage{braket}

\graphicspath{{figures_main/}{figures_sup/}}

 \usepackage[
    backend=biber,
    style=nature,
    maxbibnames=99,
  ]{biblatex}
\usepackage{times}

\newenvironment{sciabstract}{%
\begin{quote} \bf}
{\end{quote}}

\title{Tomography of Feshbach Resonance States}
\date{}
\author
{Baruch Margulis,$^{1}$ Karl P. Horn,$^{2}$ Daniel M. Reich,$^{2}$
Meenu Upadhyay,$^{3}$ Nitzan Kahn,$^{1}$ Arthur Christianen,$^{4,5}$
Ad van der Avoird,$^{5}$ Gerrit C. Groenenboom,$^{5}$ Markus Meuwly,$^{3\ast}$
Christiane P. Koch,$^{2\dagger}$ Edvardas Narevicius,$^{1,6\ddagger}$
\\
\normalsize{$^{1}$Department of Chemical and Biological Physics, Weizmann Institute of Science, 7610001 Rehovot, Israel}\\
\normalsize{$^{2}$Dahlem Center for Complex Quantum Systems and Fachbereich Physik, Freie Universität Berlin, Arnimallee 14, 14195 Berlin, Germany}\\
\normalsize{$^{3}$Department of Chemistry, University of Basel, Basel, Switzerland}\\
\normalsize{$^{4}$Max-Planck-Institut für Quantenoptik, Hans-Kopfermann-Strasse 1, D-85748 Garching, Germany}\\
\normalsize{$^{5}$Theoretical Chemistry, Institute for Molecules and Materials,
Radboud University, Heyendaalseweg 135, 6525 AJ, Nijmegen, The Netherlands}\\
\normalsize{$^{6}$Department of Physics, Technische Universit\"at, Dortmund, Germany}\\

\normalsize{$^\ast$email: m.meuwly@unibas.ch}\\
\normalsize{$^\dagger$email: christiane.koch@fu-berlin.de}\\
\normalsize{$^\ddagger$email: edvardas.narevicius@tu-dortmund.de}
}

\begin{document}

\pagenumbering{gobble}
\maketitle

\begin{sciabstract} {
Feshbach resonances are fundamental to interparticle interactions and become particularly important in cold collisions with atoms, ions, and molecules. Here we present the detection of Feshbach resonances in a benchmark system for strongly interacting and highly anisotropic collisions -- molecular hydrogen ions colliding with noble gas atoms. The collisions are launched by cold Penning ionization exclusively populating Feshbach resonances that span both short- and long-range parts of the interaction potential. We resolved all final molecular channels in a tomographic manner using ion-electron coincidence detection. We demonstrate the non-statistical nature of the final state distribution. By performing quantum scattering calculations on ab initio potential energy surfaces, we show that the isolation of the Feshbach resonance pathways reveals their distinctive fingerprints in the collision outcome.
}
\end{sciabstract}

\subsection*{Introduction}
In atomic and molecular collisions, elastic, inelastic, and reactive scattering compete with resonance pathways. In particular, Feshbach scattering resonances that arise from coupling between different degrees of freedom are instrumental for transferring energy from internal to relative motion \cite{feshbach1958unified,fano1961effects}. 
Feshbach resonances have been observed in atom-atom \cite{frisch2014quantum,barbe2018observation,chin2010feshbach}, atom-ion \cite{weckesser2021observation}, atom-molecule \cite{son2022control,de2020imaging,yang2019observation}, bimolecular \cite{park2023feshbach,chefdeville2013observation} and field dressed bimolecular collisions\cite{chen2023field} by tuning either the collision energy or the magnetic field strength in the case of trapped ultra-cold particles. 
Alternatively, a single resonance path can be selected by spectroscopic excitation, as in predissociation \cite{crim1990state,miller1988vibrational,rohrbacher2000dynamics,foley2021orbiting} or electron photodetachment \cite{kim2015spectroscopic,otto2014imaging}. However, one is then limited to the Franck-Condon region close to the equilibrium geometry which represents a rather small part of the overall phase space.
Here, we combined the advantage of ultracold molecular collisions exploring a large part of the phase space with the preparation of isolated resonances that allowed us to follow collision pathways. This was achieved by populating vibrational Feshbach resonances at long range, just below the threshold. In contrast to predissociation and similarly to full collisions, the colliding particles possess a set of well-defined unperturbed internal quantum numbers. 
As in reactive bi-alkali systems \cite{son2022control,yang2019observation,ni2008high} we explored the collision regime where the short range intermolecular interaction is strong, leading to mixing between tens of rovibrational quantum states coupled by anisotropic interactions. 

Our method is complementary to and expands the capabilities of the electron photodetachment spectroscopy (EPDS)\cite{kim2015spectroscopic,otto2014imaging}(see, SM). In EPDS, the electron is stripped away from an anion by interaction with laser radiation. The electron kinetic energy distribution contains information about the energy levels of the neutral product. However, the detection of subsequent dynamics is challenging. In our method, the starting point is a cold collision between an electronically excited atom and a neutral molecule. The electron is emitted during a charge transfer process resulting in ionic molecular products with a well defined set of internal quantum numbers. Due to the low collision energy, we populated a narrow momentum band during the ionization step. In such a case, the distribution of initial and final molecular ion states can be simply detected using energy conservation. 
The kinetic energy distribution of the molecular ions provides additional information which encodes the final quantum state distribution resulting from the decay of Feshbach resonances populated upon the ionization step. 
The non-selective nature of the ionization process leads to the formation of molecular ions in many vibrational states. As a result, the ionic kinetic energy distribution contains contributions from all the accessible vibrational branches of Feshbach resonances. Crucially, we disentangled the contribution of individual initial molecular ionic states using ion-electron coincidence detection. This scheme allowed us to simultaneously detect all the possible quantum states in a tomographic manner. Our results show that each Feshbach resonance has a distinctive signature reflected in the final quantum state distribution. In a similarly strongly interacting system it has been shown \cite{liu2021precision} that the final state distribution follows statistical arguments. In earlier reactive scattering experiments where Feshbach resonances were detected \cite{skodje2000resonance,qiu2006observation}, the measured final state distribution was found to be insensitive to the collision energy up to 180K and did not depend on the total angular momentum either\cite{wang2014isotope}. By contrast, in our case the quantum signatures of vibrational resonances are sensitive to the details of the interaction. We confirmed this fact by carrying out full \textit{ab initio} based quantum scattering calculations for the benchmark system of H$_2^+$ colliding with a noble gas atom. Our work shows that quantum signatures can be observed even in strongly interacting and anisotropic systems if the resonance pathway can be isolated.

\subsection*{The Method}
\begin{figure}[t]
    \centering
    \includegraphics[width=1\textwidth]{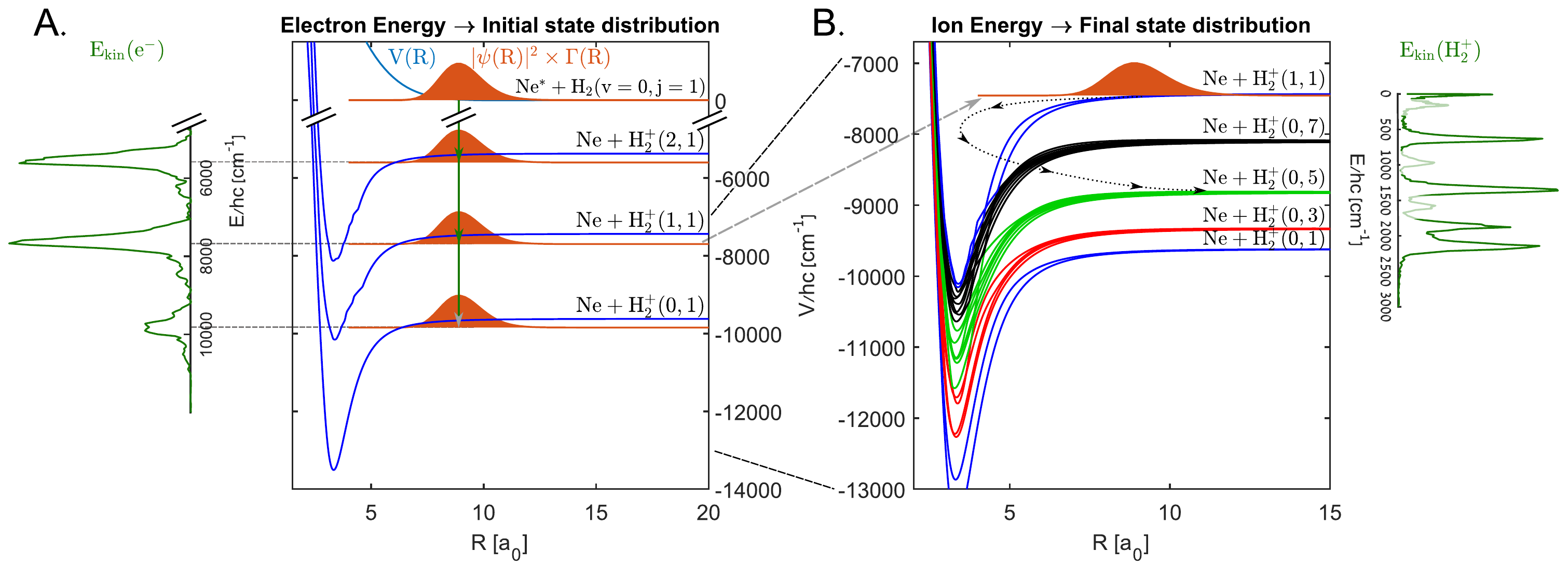}
    \caption{Adiabatic potential energy curves and wave functions describing Penning collisions at total angular momentum state $J=6$ and partial wave $l=5$ between metastable neon and ortho-H$_2$ in its ground rovibrational state. Each adiabatic curve corresponds to the vibrational and rotational quantum numbers, $(v,j)$, of $\mathrm{H_2^+}$. $\mathrm{R}$ is the intermolecular separation, $a_0$ is the bohr radius. Same color curves represent adiabatic curves with different partial wave quantum number, $l$, but with the same molecular rotation quantum number, $j$. The left side (A.) of the figure represents the ionization step. The green curve presents the measured kinetic energy of electrons, providing the initial state distribution of the Feshbach states. The energy position of the projected states are set relative to the center of the measured peaks.
    The right side (B.) of the figure represents the decay of Feshbach states formed upon ionization.
     The measured ionic kinetic energy distribution (green curve) corresponds to the final state distribution. The contribution of para-H$_2$ to the kinetic energy distribution is marked in light green. The energy position of the projected state is set by the measured resonance energy.}
    \label{dynamics}
\end{figure}

We present a schematic description of our approach in Fig.~\ref{dynamics}. The starting point is the collision between a metastable noble gas atom and a ground state molecule, which leads to Penning ionization. The ionization is a sudden, non-adiabatic transition of the initial neutral state onto the manifold of ionic states\cite{miller1970theory}. It can be described by projecting the scattering wave function of the neutrals $\mathrm{\psi(R)}$, scaled by the ionization probability $\mathrm{\Gamma(R)}$, onto the full set of states describing the molecular ion/neutral atom interactions. Because ionization occurs at large intermolecular distances, the initial state is projected in the region where the long-range part of the intermolecular potential dominates. As a result, the Feshbach states are characterized by the free molecule basis with no change in the rotational quantum number, $j$, of the molecule.
The populated states differ asymptotically by the vibrational energy of the molecular ion. Due to the electron's small mass, it carries away most of the excess energy, generated at the ionization step. Therefore, the initial vibrational state, $v$, of the molecular ion is encoded in the electron kinetic energy. The energy distribution of electrons represents the vibrational spectrum of the molecular ion with near Franck-Condon probabilities\cite{siska1993molecular,tanteri2021study}.
In Fig.~\ref{dynamics}B we show the resulting dynamics of such a Feshbach resonance state where a single quantum of vibrational excitation is converted into a combination of kinetic and rotational energy. Here, the closed channel is the vibrationally excited state, whereas the open channels are the vibrationally ground, rotationally excited states. The final quantum state distribution is reflected in the velocity distribution of the molecular ion. The electron-ion coincidence measurement completes the quantum state-to-state mapping because every detected electron-ion pair is generated during the same collision event.

\begin{figure}[t]
    \centering
        \includegraphics[width=0.9\textwidth]{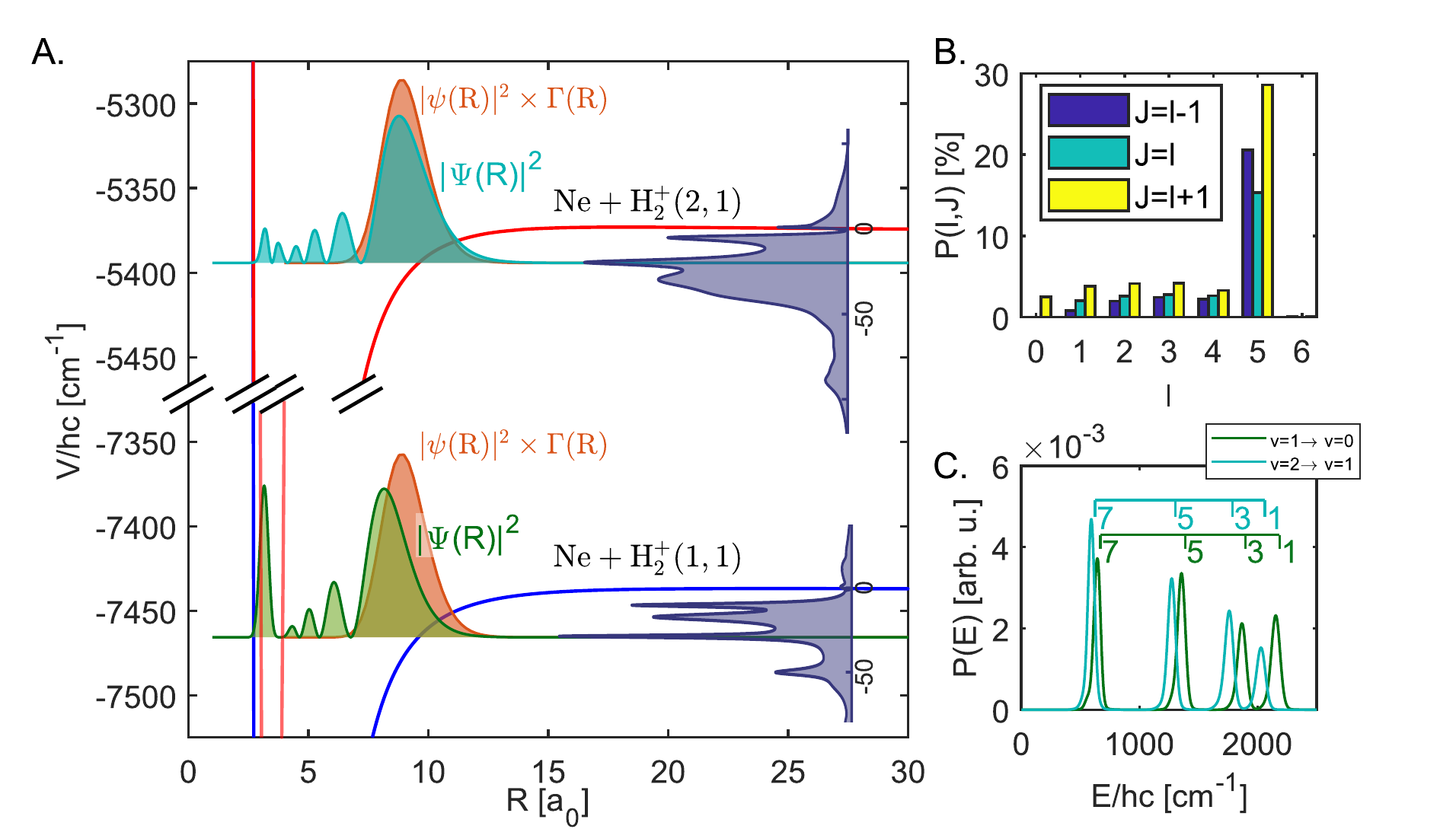} %6 - without initial wave packet
    \caption{Computational characteristics of Feshbach resonance states. A. Zoomed in view of adiabatic PES together with corresponding radial wave functions $\mathrm{\Psi(R)}$. Shown on the right side of the figure are computed overlap plots for $J=6$ and initial $l=5$ which represent the energy distribution of populated Feshbach states. The presented ionic energy eigenstates have the largest overlap with the initial neutral state describing the Penning collisions ($\mathrm{|\psi(R)|^2\times \Gamma(R)}$). B. Relative contribution of each angular momentum state of collision for $\mathrm{Ne(^3P)-ortho-H_2}$ collisions considering the experimental collision energy and spread. C. Computed ionic energy distributions representing final state distribution corresponding to initial vibrational state $v=2$ and $v=1$. The threshold energies are labeled by the final rotational state of free ortho-H$_2^+$ ions.  The distributions are normalized to unity.}
    \label{F_state}
\end{figure}

The ionization step is governed by a charge transfer process that depends exponentially on the intermolecular distance. As such, the part of the scattering wave function that undergoes ionization is strongly localized in the vicinity of the classical turning point of the neutral potential energy surface (Fig.~\ref{dynamics}A). Such a strong localization of the initial wavefunction is essential for the resolution of our measurement since it determines the energy spread in the distribution of populated ionic states. After ionization, the initial scattering wavefunction spans a narrow, 50 $\mathrm{cm^{-1}}$ wide energy band located 30 $\mathrm{cm^{-1}}$ away from the dissociation threshold of the molecular-ion/neutral-atom interaction potential energy surface. This energy window constitutes a small fraction of the total interaction strength reflected by a deep potential well of 4450 $\mathrm{cm^{-1}}$ in the case of $\mathrm{Ne-H_2^+}$. This energy window contains only a few Feshbach resonances, associated to triatomic rovibrational states (see Fig.~\ref{F_state} A. for two initial vibrational levels of the molecular hydrogen ion, $v=2$ and $v=1$). Moreover, the total angular momentum and parity of the Feshbach resonance are set during the Penning ionization step. By tuning the collision energy of the neutral collision system to match a shape resonance, we were able to control the total angular momentum distribution which peaked around the resonant partial wave\cite{henson2012observation,klein2017directly}.
Importantly, the near-threshold Feshbach resonance wavefunctions are sensitive to both the short and the long range parts of the interaction potential. While in the short range, all of the degrees of freedom are strongly coupled by the anisotropy of the potential energy surface (PES), at large separation, the resonance wavefunction is nearly separable into the molecular ion and neutral atom. Whereas the strength of the leading term of the anisotropic interaction is  1300$\,\mathrm{cm^{-1}}$ at an intermolecular separation of 2$a_0$, it falls to 1 $\,\mathrm{cm^{-1}}$  at 10$a_0$. We demonstrate that each Feshbach resonance state has a distinctive signature in the final state distribution, changing with the Feshbach state energy, the total angular momentum, and the vibrational quantum number, cf. Fig~\ref{F_state}C.

\subsection*{Results and Discussion}
\begin{figure}[t]
    \centering
    \includegraphics[width=1\textwidth]{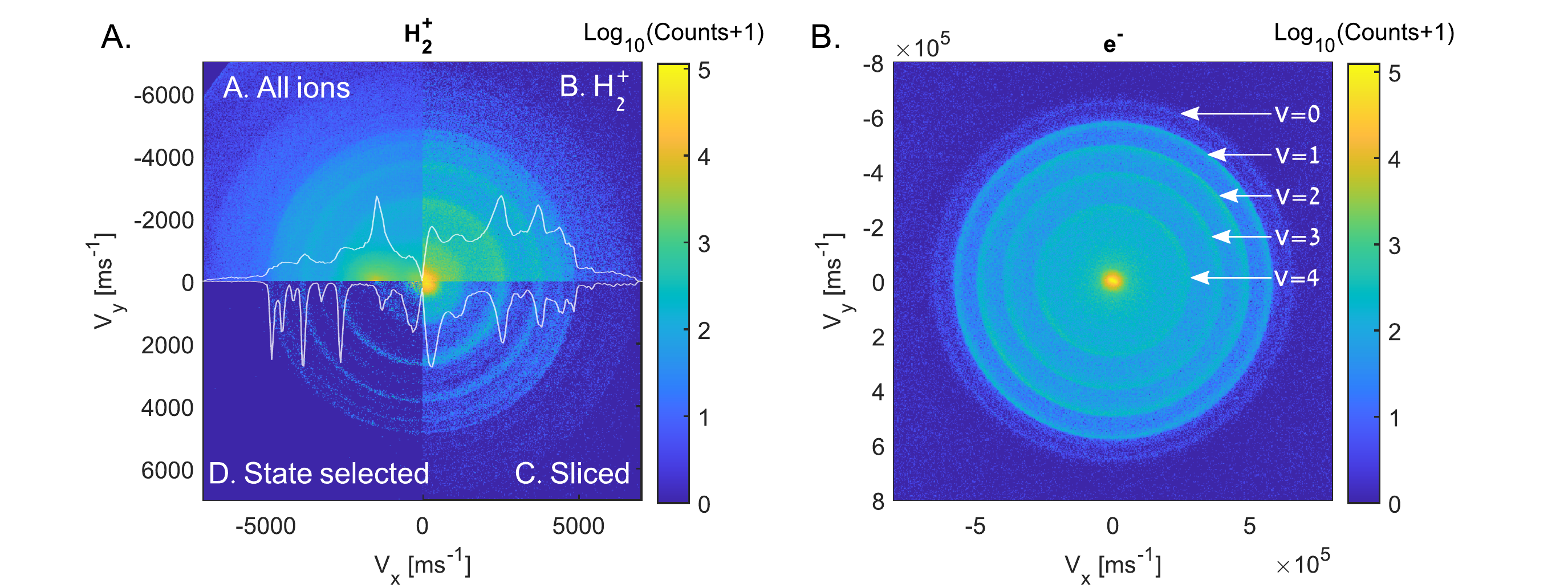} 
    \caption{VMI images of ions and coincidence electrons. A. going clockwise from the top left corner of the image, the first quadrant presents the overall accumulated ionic VMI image which includes all ionic masses, the second quadrant shows the VMI image of H$_2^+$ ions alone obtained using electron-ion coincidence, the third quadrant presents a sliced image using ion time-of-flight information and the last quadrant presents the sliced VMI image of ions related to initial vibrational state $v=1$ using ion-electron correlations. Angle-integrated velocity distributions are presented as white curves at each quadrant. B. VMI image of coincidence electrons corresponding to H$_2^+$.}
    \label{sl_VMI_NeH2} 
\end{figure}

We studied two benchmark molecular ion collision systems, $\mathrm{Ne-H_2^+}$ and $\mathrm{He-H_2^+}$ by measuring correlated energy distributions of products of $\mathrm{Ne^{*}-H_2}$ and $\mathrm{He^{*}-H_2}$ PI collisions. The experimental setup is described in detail elsewhere\cite{margulis2020direct}. In short, we generated two supersonic beams of metastable noble gas atoms and ground state molecules using two Even-Lavie\cite{even2000cooling} valves which were positioned at a relative angle of 4.5 degrees. We excited the noble gas atoms to a metastable state by a dielectric discharge which was located directly after the valve orifice\cite{luria2009dielectric}. The beam mean velocity was determined by the valve temperature and the gas mixture composition.
We performed Penning ionization collisions at an energy that matched the position of one of the shape resonances that occur at a few Kelvin\cite{henson2012observation}. We first located the shape resonance positions by a complementary measurement using the merged beam approach (see, SM). We present the angular momentum state distribution for Ne$^*$ - ortho-H$_2$ and Ne$^*$-para-H$_2$ and He$^*$-H$_2$ collisions in Fig.~\ref{F_state}B and  Fig.~S2 %\ref{w_perj}
 respectively.
For $\mathrm{Ne^*+H_2}$, we chose a collision  energy matching the position of the $l=5$ shape resonance. As a result, scattering states with total angular momenta $J=4,5,6$ had the highest contribution.  

We detected the product ion and electron velocities using a Coincidence Double Velocity Map Imaging (CDVMI) apparatus\cite{margulis2020direct}. Coincidence detection was the most important aspect of our experiment. The correlation between ions and electrons allowed us to extract the mass, identify the initial state, and determine the velocity magnitude distribution instead of just the projection. It also allowed us to correct the ionic VMI images for electron recoil. Such a correction was essential for He-H$_2$ where the electron recoil blurred the ionic VMI image by about 100m/s .

We present the ion and electron VMI images as measured in coincidence during the  $\mathrm{Ne^*-H_2}$ collisions in  Fig~\ref{sl_VMI_NeH2}.
The ion VMI image is divided into quadrants which represent the various steps in the analysis of the raw, accumulated data. 
We started with the raw accumulated VMI image which includes all the ionic products in Fig.~\ref{sl_VMI_NeH2}A. We used TOF information to obtain the VMI image of H$_2^+$ ions alone, Fig.~\ref{sl_VMI_NeH2}B. During the next step we selected the central part of H$_2^+$ TOF distribution in order to time-slice the VMI image (Fig.~\ref{sl_VMI_NeH2}C). This procedure enabled us to construct the velocity magnitude distribution by angle integration of a sliced image without the need of inverse Abel transformation based techniques. The mass selected and time-sliced ionic VMI image consists of a series of concentric rings indicating a set of discrete final quantum state channels where the internal vibrational molecular ion energy is converted into a combination of the final rovibrational state and translational energy. However, this image contains information from all the five initial vibrational states of the H$_2^+$ molecular ion that are populated during the quenching process. We constructed the ionic VMI image which corresponds to an individual initial quantum state of H$_2^+$ by post-selecting only those ions which correlate with the electrons within a certain kinetic energy window. Our tomography procedure required eliminating the effect of high energy electrons on rings associated with the lower energy electrons. We have developed a 'peeling' algorithm that corrects an individual ionic image due to the overlap of its coincidence electrons with higher energy electrons. A detailed description of the 'peeling' algorithm can be found in the SM.

%%%%%%%%%%%%%%%%%%%%%%%% Final state distribution %%%%%%%%%%%%%%%%%%%

\begin{figure}[t!]
    \centering
     \includegraphics[width=0.9\textwidth]{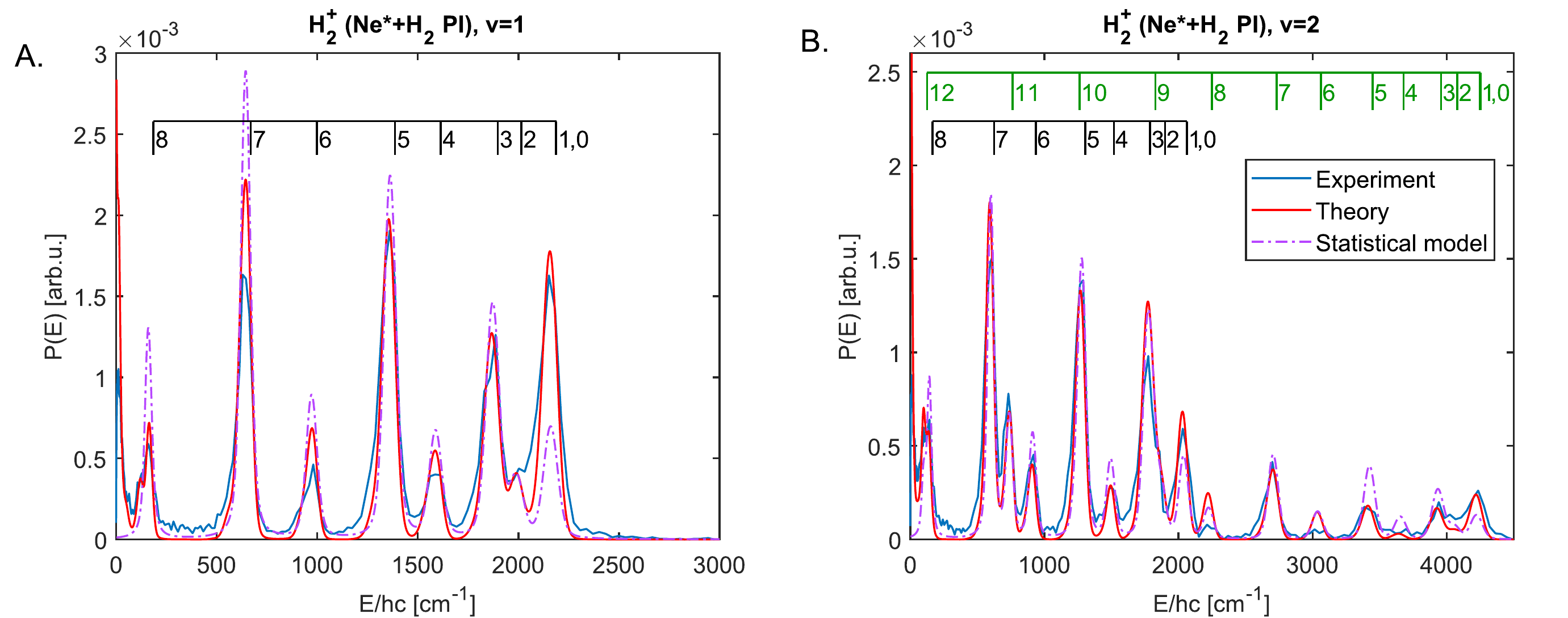}
     \includegraphics[width=0.9\textwidth]{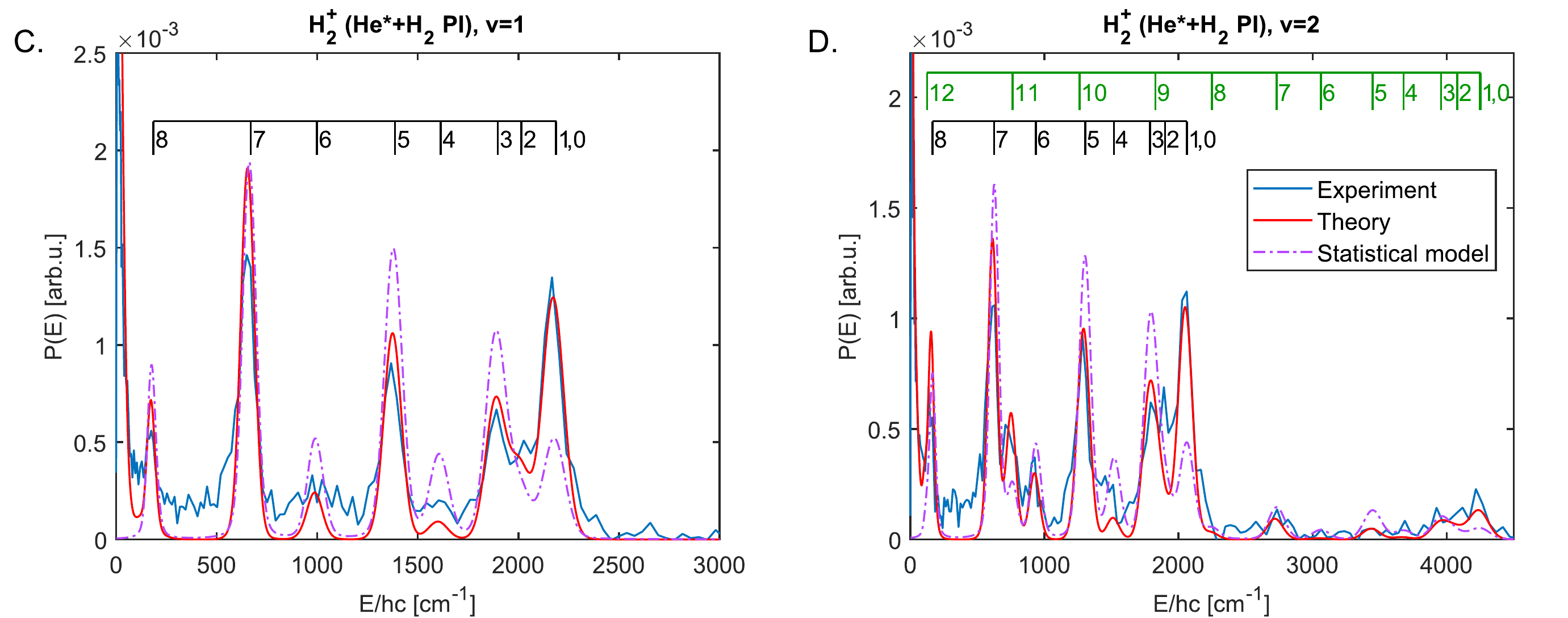}
    \caption{Experimental (blue) and theoretical (red) kinetic energy distributions showing the decay of Feshbach states into various continuum states for $\mathrm{He-H_2^+}$ (bottom) and $\mathrm{Ne-H_2^+}$(top). Left column is for initial $v=1$, right column is for $v=2$. Theoretical curves are convoluted by the experimental resolution. Distributions based on the statistical model (see text) are the pink dashed curves.}
    \label{tom_NeH2}
\end{figure}

We obtained the energy distribution of the molecular ions by angular integration of the sliced, initial-state-selected VMI images (blue curves in Fig \ref{tom_NeH2}). The energy scale corresponds to the kinetic energy release, E, which was partitioned between the molecular ion and the neutral atom according to momentum conservation. We observed sharp peaks indicating the quenching of Feshbach resonances to different continuum states which were labeled by the final quantum state, ($v',j')$, of the free molecular ion. The final state distribution thus represents the observation of the projection of the Feshbach resonance onto the continuum basis set which we refer to as tomography. The threshold energies are depicted by black and green lines for $\Delta v=1$ and $\Delta v=2$ transitions, respectively. Each threshold energy is labeled by the final rotational state, $j'$, of the free H$_2^+$ ion.
We observed a higher probability of odd final rotational states due to the spin statistics of ortho vs para hydrogen spin isomers with the 3:1 statistical population ratio.
In addition to the resonance state tomography, we obtained the resonance energy which is reflected by the shift of all peaks relative to the threshold energies. For the $v=1$ state, we found the mean resonance energies to be $-28.9 \pm 3.6$ cm$^{-1}$ for $\mathrm{Ne-H_2^+}$  and $-20.8 \pm 3.1$ cm$^{-1}$ for $\mathrm{He-H_2^+}$. All resonance energies are reported relative to the position of the noble gas atom and rotational ground state molecule dissociation threshold.

To confirm the nature of the resonances and to provide additional characterization, quantum wave packet calculations were carried out on an available full configuration interaction (FCI) PES for $\mathrm{He-H_2^+}$\cite{koner2019heh2+pes} and a new PES for $\mathrm{Ne-H_2^+}$. The new $^2$A' $\mathrm{Ne-H_2^+}$ PES utilizing 38200 reference energies was determined at the  CCSD(T)/aug-cc-pV5z level of theory. Full coupled channel calculations of the $\mathrm{He-H_2^+}$ and $\mathrm{Ne-H_2^+}$ cross sections were performed with converged bases with rotational states going up to and including $j=25$ and vibrational states going up to and including $v=5$ respectively, though for $\mathrm{He-H_2^+}$  fewer rotational states were required to obtain convergence.
The numerical results were convoluted with the experimental uncertainty and are in excellent agreement with the experimental data (blue vs. red traces in Fig.~\ref{tom_NeH2}). Together with earlier work on the near-dissociative states of $\mathrm{He-H_2^+}$~\cite{koner2019heh2+pes}, the present agreement between experiment and theory points to near-spectroscopic accuracy of the FCI PES. Similarly, for Ne--H$_2^+$, the agreement between computations and experiment is very encouraging. Further improvement could be obtained from morphing the underlying PES~\cite{meuwly1999morphing,bowman1991simple}.

%%%%%%Statistical model%%%%%%%%%
Due to the strong and highly anisotropic intermolecular interaction one could expect a statistical description to be sufficient to describe the final state distribution. In fact, both interaction strength and anisotropy are comparable to the case of KRb+KRb bi-molecular reactions~\cite{liu2021precision}. In the bi-molecular case, deviations from the statistical model were found at the limit of the reaction exothermicity where the centrifugal barrier hinders the reaction rate. In our case, the level of initial excitation relative to the centrifugal barrier height in the exit channel (Fig.\ref{dynamics}B) allows all vibrational and rotational transitions to occur. In a purely statistical approach, the decay probability is proportional to the number of energetically accessible exit channels~\cite{pechukas1966statistical,liu2021precision}.
The level of degeneracy in each continuum state is set by the initial total angular momentum $J$ and by parity $(-1)^{j+l}$, which are conserved quantities. For a final rotational state $j'$, the number of exit channels is determined by the number of available partial waves which fulfill the relation $J=|j'-l'|,|j'-l'|+1,\ldots,j'+l'$ and maintain the right parity. We have constructed a statistical model where we assumed a single Feshbach resonance, at position and width as obtained by exact calculations, where the decay probability to each final rovibrational state is given by the number of allowed partial waves. The resulting energy distribution was convoluted by the experimental resolution and shown as pink dashed lines in Fig \ref{tom_NeH2}. Comparing the results of this procedure to the experimental data shows clear deviations from the statistical model. A quantitative comparison of the results of the statistical model and exact calculations to the experimental data is shown in Fig.~S10. %\ref{comp_st_exp}.
In particular, the statistical model underestimates the decay to states with low final $j'$ and overestimates the decay to states with high final $j'$. This observation suggests that certain transition probabilities are preferred rather than statistical. 

\subsection*{Outlook}
Although we probed a collision system with strong interactions (charge - induced dipole) and high anisotropy, the Feshbach resonance states have a unique quantum signature in the final state distribution. This property may be useful for control over the final state distribution by manipulating the quantum state of the Feshbach state itself. One way to do so is by tuning the total angular momentum which is difficult to control. However, an effective way to select states with a certain amount of  total angular momentum is to leverage different shape resonances of the neutral collision complex. This could be done using the merged beam approach where collision dynamics could be tuned down to a single quantum of angular momentum~\cite{margulis2022observation}. Performing coincidence experiments in the p-wave limit will demonstrate the control of the final state distribution. Moreover, due to the reduction in participating quantum states, individual Feshbach states corresponding to the triatomic rovibrational spectrum, could be resolved.

\newpage
%\bibliography{refs}
%\bibliographystyle{Science}

  \printbibliography
  
\newpage

\section*{Acknowledgements}
We thank F. Gianturco for sharing preliminary cross sections that allowed us to test our own wavepacket simulation code.
\textbf{Funding:} Financial support from the European Research Council Advanced Grant (E.N.), the NCCR MUST and the University of Basel (M.M.), and the Deutsche Forschungsgemeinschaft (Project "Efficient Quantum Control of Molecular Rotations", Grant No. 505622963, C.P.K.) is gratefully acknowledged.
\textbf{Author contribution:} B.M., N.K. and E.N. designed, constructed, conducted the experiment and analyzed the results. M. U. and M. M. designed, computed and tested the potential energy surfaces. K.P.H. and D.M.R. wrote the scattering software with the help of A.C., A.v.d.A, and G.C.G. and calculated theoretical cross sections which were analyzed by K.P.H., D.M.R., A.C., and C.P.K. All authors contributed to writing and revising the manuscript.
\textbf{Competing interests:} The authors declare that they have no competing interests. \textbf{Data and materials availability:} All data underlying the figures are deposited at Zenodo \cite{Zenodo}. The information for the potential energy surfaces is available at https://github.com/MMunibas/rgh2.

\subsection*{List of supplementary materials:}
Materials and Methods.\\
Supplementary text. \\
Figs S1 to S10. \\
Table S1. \\
References 38-51.\\

%\nocite{paliwal2021determining,pawlak2015adiabatic,pawlak2017adiabatic,Johnson_1973,ashfold2006imaging,townsend2003direct,dick2014inverting,unke2017toolkit,werner2020molpro,gammie2002microwave,carrington1996observation,pack1973space,johnson1978renormalized,gadea1997nonradiative}

\newpage

%%%The following handles adding "S" in front of figures in SI _and_ in the main text...
\renewcommand{\thetable}{S\arabic{table}}
\renewcommand{\thefigure}{S\arabic{figure}}
\setcounter{figure}{0}    

\title{Supplementary Materials for\\
    \textbf{Tomography of Feshbach resonance states}}
\date{}
\renewcommand{\Authands}{ and } 

%\begin{document}

\newcounter{Sequ}
\newenvironment{SEquation}
  {\stepcounter{Sequ}%
    \addtocounter{equation}{-1}%
    \renewcommand\theequation{S\arabic{Sequ}}\equation}
  {\endequation}

\newenvironment{SEqnarray}
    {\stepcounter{Sequ}
    \addtocounter{equation}{-1}
    \renewcommand\theequation{S\arabic{Sequ}}\eqnarray}    
    {\endeqnarray}

\begin{center}
\end{center}

{\let\newpage\relax\maketitle}

\textbf{\noindent The PDF file includes:}\\
\indent \indent Materials and Methods.\\
\indent \indent Supplementary text. \\
\indent \indent Figs S1 to S10. \\
\indent \indent Table S1. \\

\newpage

\subsection*{Materials and Methods}
\underline{Supersonic beams properties}\\

The supersonic beams properties are presented in Table~\ref{tab:my_label}. Metastable beam profiles were acquired using a microchannel plate (MCP), positioned in line with the beam travel direction. Neutral beam velocities were deduced from the delay between the beams, beam spreads were calculated using a delay scan between the beams. The collision energy was calculated according to $E=\frac{1}{2}\mu \left( V_1^2+V_2^2-2V_1V_2\cos(\theta) \right)$, where $\theta=4.5^0$ is the angle between the valves and $\mu$ the reduced mass. The spread in collision energy was obtained by simulating the beam trajectories, taking into account the experimental mean velocities and spreads.

\begin{center}
\begin{table}[tb!]
    \centering
\resizebox{\columnwidth}{!}{\begin{tabular}{ccccccccc} 
 \hline
 System & $T_1$ [K] & $V_1$ [ms$^{-1}$] & $dv_1$ [ms$^{-1}$] & $T_2$ [K] & $V_2$ [ms$^{-1}$] & $dV_2$ [ms$^{-1}$] & $E/k_B$ [K] & $dE/k_B$ [K] \\ [0.5ex] 
 \hline\hline
 He$^*$+H$_2$ & 170 & 1346 & 22 & 139 & 1245 & 38 & 2.25 & 0.67\\ 
 \hline
 Ne$^*$+H$_2$ & 205 & 659 & 14 & 140 & 811 & 21 & 2.93 & 0.44   \\
 \hline
 \end{tabular}}
    \caption{Supersonic beams properties. $T_1$, $V_1$, $dV_1$ are the valve temperature, mean velocity, and one standard deviation of velocity spread for the metastable beam. $E$ and $dE$ are the collision energy and spread accordingly. The composition of the molecular beams in He$^*$+H$_2$ system: 10\% Ne in H$_2$, Ne$^*$+H$_2$ system: 50\% Ne in H$_2$.
 }
    \label{tab:my_label}
\end{table}
\end{center}

\noindent \underline{Collisions in shape resonance conditions}\\

\begin{figure}[tb]
    \centering
    \includegraphics[width=0.65\textwidth]{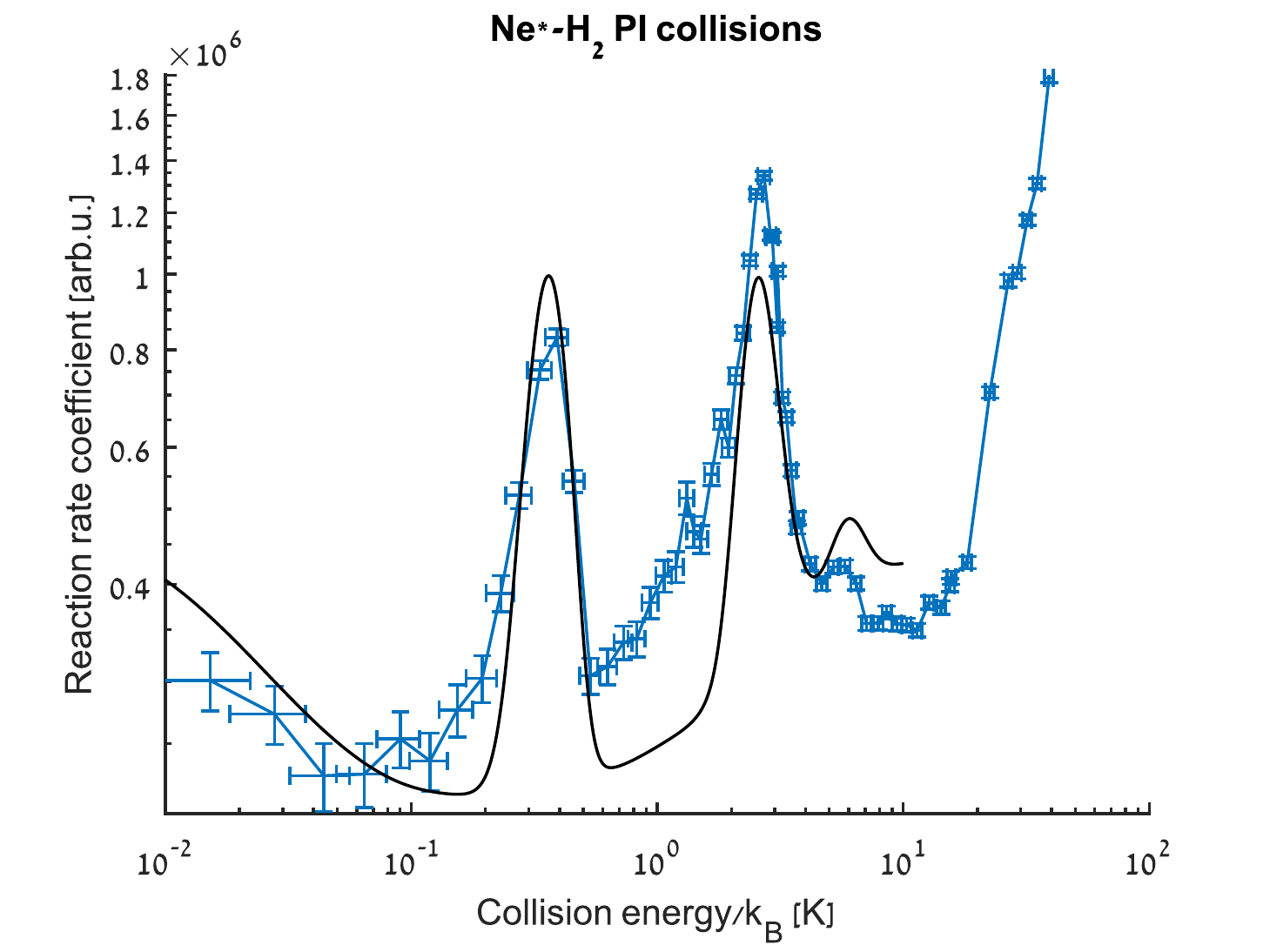}
    \caption{Reaction rate coefficients as a function of the collision energy for PI collisions of Ne$(^3P)$+H$_2$. Blue crosses indicate experimental data with error bars. The horizontal error bars represent $\pm$1 STD of statistical error. The vertical error bars represent $\pm$1 STD uncertainty in the collision energy. The black line is the theoretical curve which is convoluted by the experimental spread in collision energy. Peaks at collision energies of $k_B\times$0.4K and $k_B\times$2.7K are associated to partial wave resonances $l=4$ and $l=5$ respectively.}
    \label{NeH2HDRR}
\end{figure}

\begin{figure}[tb] 
    \centering
    \includegraphics[width=0.45\textwidth]{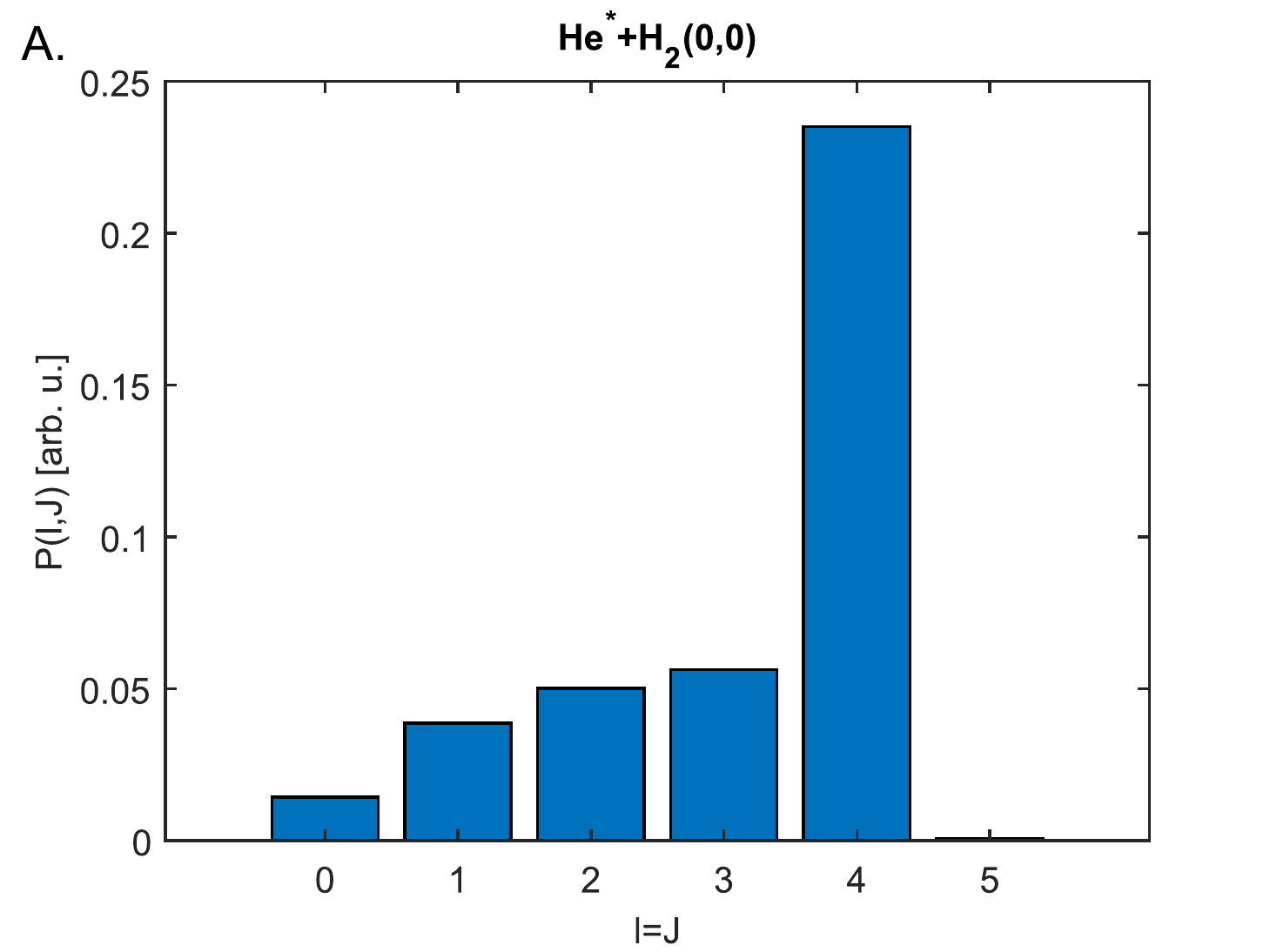}
    \includegraphics[width=0.45\textwidth]{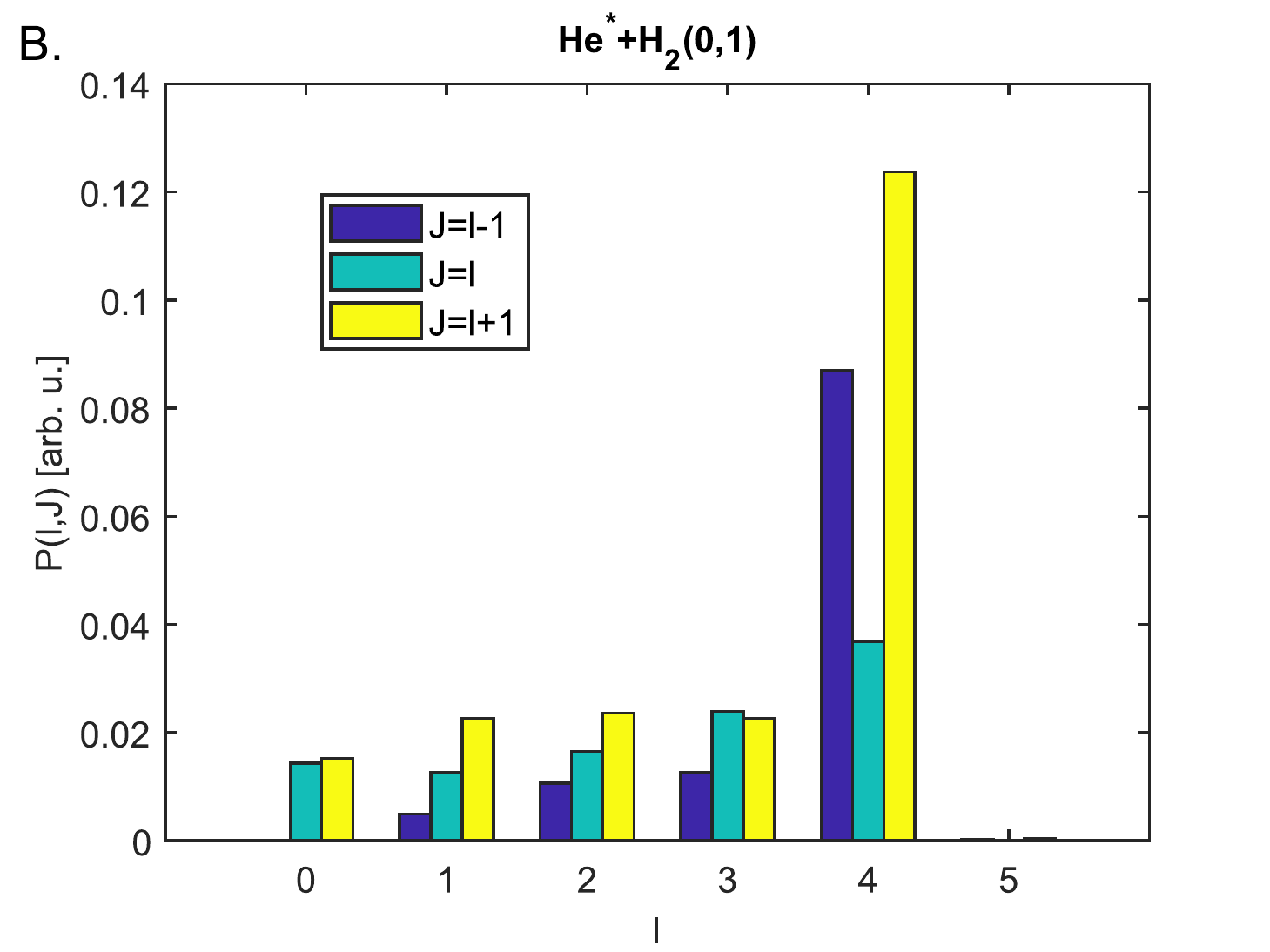} 
    \includegraphics[width=0.45\textwidth]{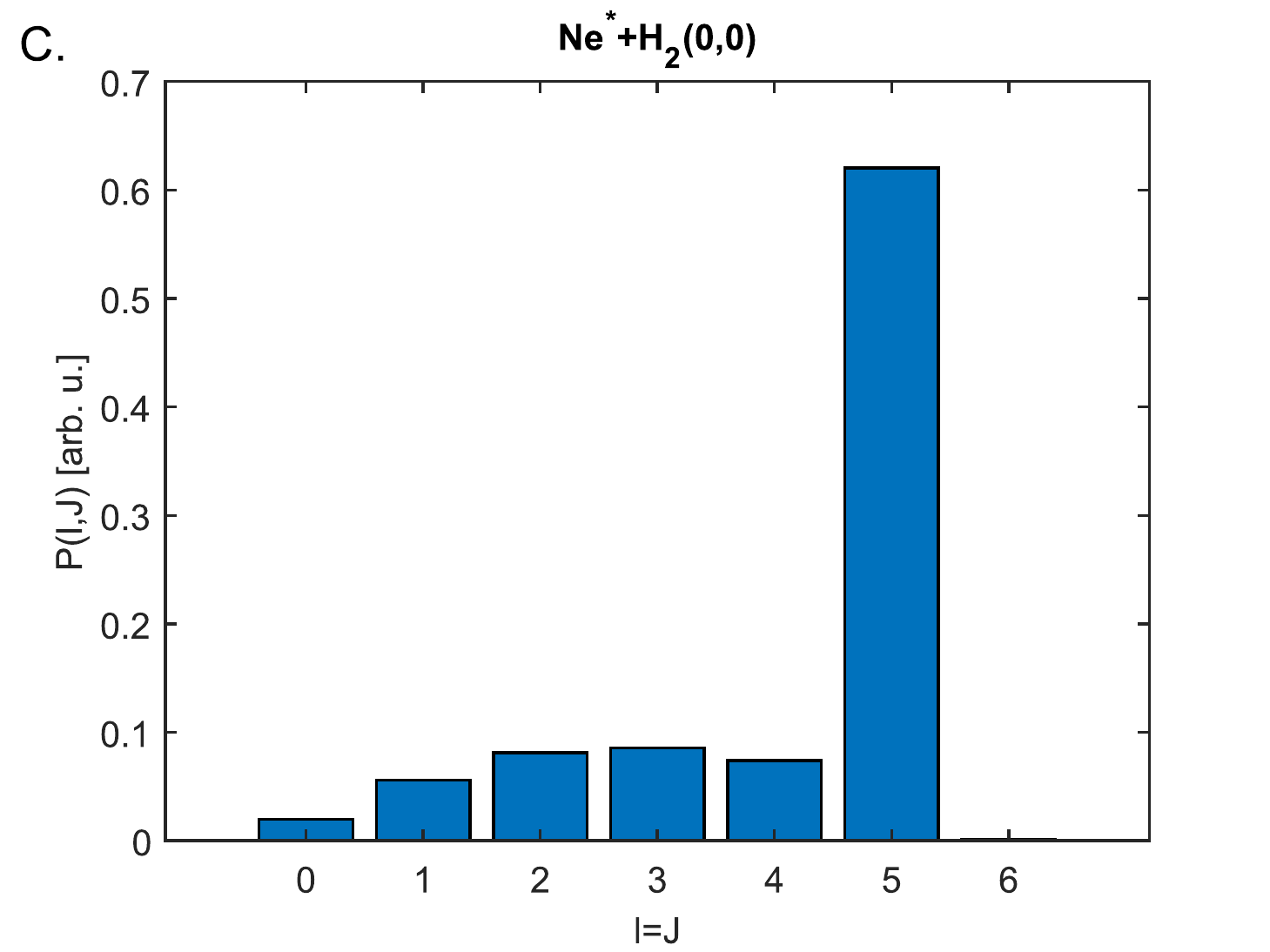} 
    \caption{Contribution of each state of total angular momentum and parity to the ionization process, calculated for the experimental collision energies and spreads (table S1).
    } 
    \label{w_perj}
\end{figure}

Performing collisions where only a few quantum states describe the interaction is essential for state-to-state resolution. For this reason, we have chosen to perform Penning ionization (PI) collisions at a shape resonance. Shape resonances are long-lived quasi-bound states, formed by tunneling through a centrifugal barrier. As such they are associated with a specific angular momentum $l$ of the collision complex. 
The resonance energies were found by an independent measurement using a merged beams experimental setup which is described elsewhere~\cite{paliwal2021determining,margulis2022observation}. Reaction rate coefficients as functions of the collision energy are presented in Fig.~\ref{NeH2HDRR} for $\mathrm{Ne^*(^3P)-H_2}$, whereas the experimental data for $\mathrm{He(^3S)-H_2}$ collisions is found elsewhere~\cite{klein2017directly}. The rate coefficients were calculated as the ratio between counts of ions, per collision energy, to the intensity of the reactant beams. The horizontal error bars represent statistical errors, estimated as the square root of the number of counts. The vertical error bars represent the uncertainty in the collision energy, calculated from the uncertainty in the mean velocity of individual beams. The latter was derived from the uncertainty in travel distance ($\sim$10 mm) and time ($\sim$20$\mu s$).
The reaction rate coefficient peaks at collision energies associated with shape resonances. Theoretical curves were calculated using the adiabatic approach~\cite{pawlak2015adiabatic,pawlak2017adiabatic} where the scattering wave function is propagated from 7$\,a_0$ to 500$\,a_0$ (with $a_0$ the bohr radius) applying the log-derivative method~\cite{Johnson_1973}. Details regarding the used potential energy surfaces can be found elsewhere~\cite{margulis2022observation,klein2017directly}.
A necessary input for the theoretical calculations is the ionization probability per initial state of the collision, labeled by partial wave, $l$, molecular rotation, $j$ and the total angular momentum, $J=l+j$ which is a conserved quantity. $j$ is either 0, for parahydrogen, or 1, for orthohydrogen. We have obtained the contribution of each initial state by integrating the theoretical reaction rate coefficients in the window of experimental collision energies. The results, shown in the $J,l$-basis are presented in Fig~\ref{w_perj} and in Fig~\ref{F_state}. For collisions with orthohydrogen, three $J$ states with $J=l_\mathrm{res}-1:l_\mathrm{res}+1$ dominate the reaction.\\

\noindent \underline{Data analysis}\\

\noindent \underline{Mass spectrum}\\

\begin{figure}[tb!]
    \centering
    \includegraphics[width=0.48\textwidth]{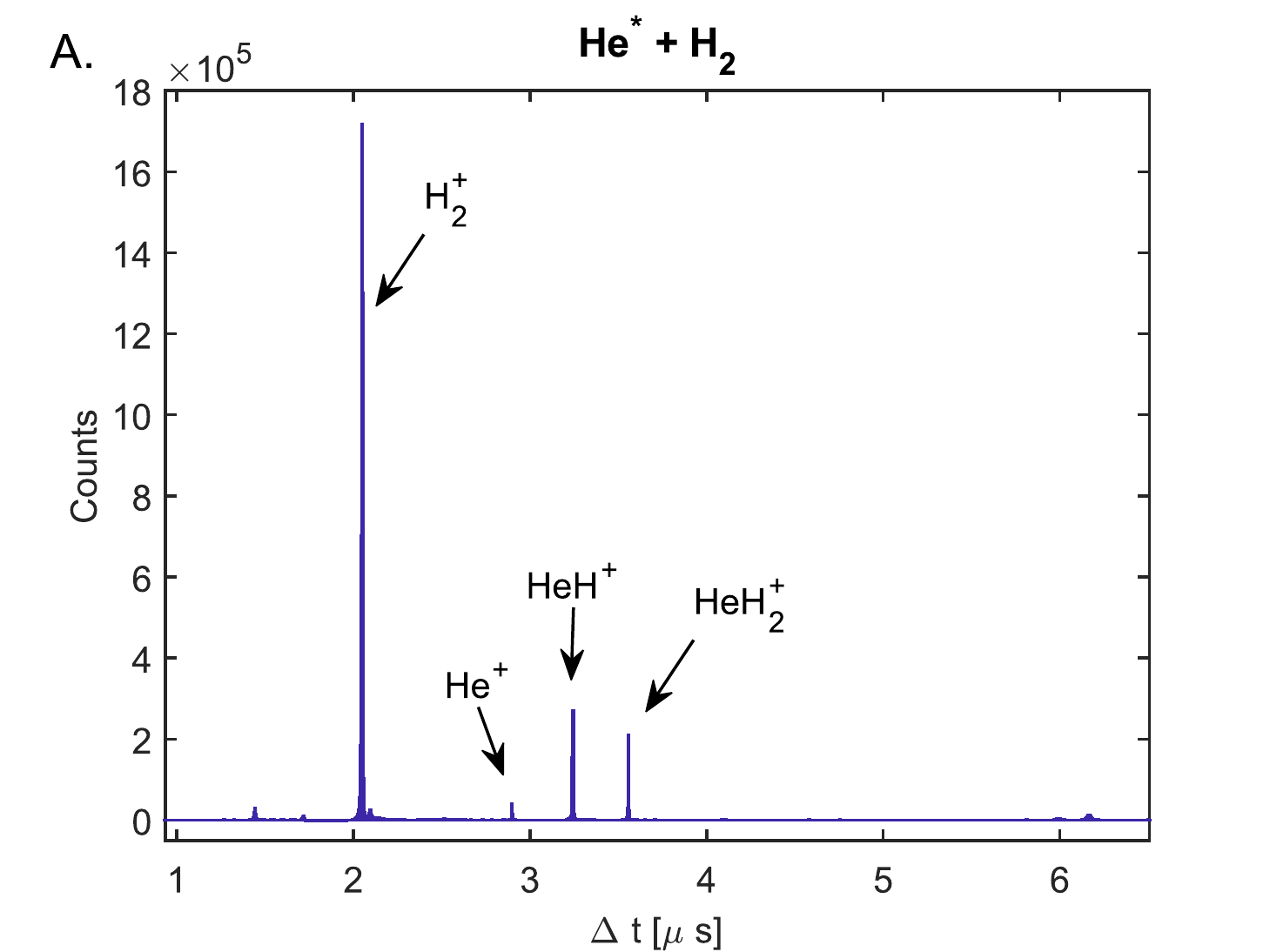} 
    \includegraphics[width=0.48\textwidth]{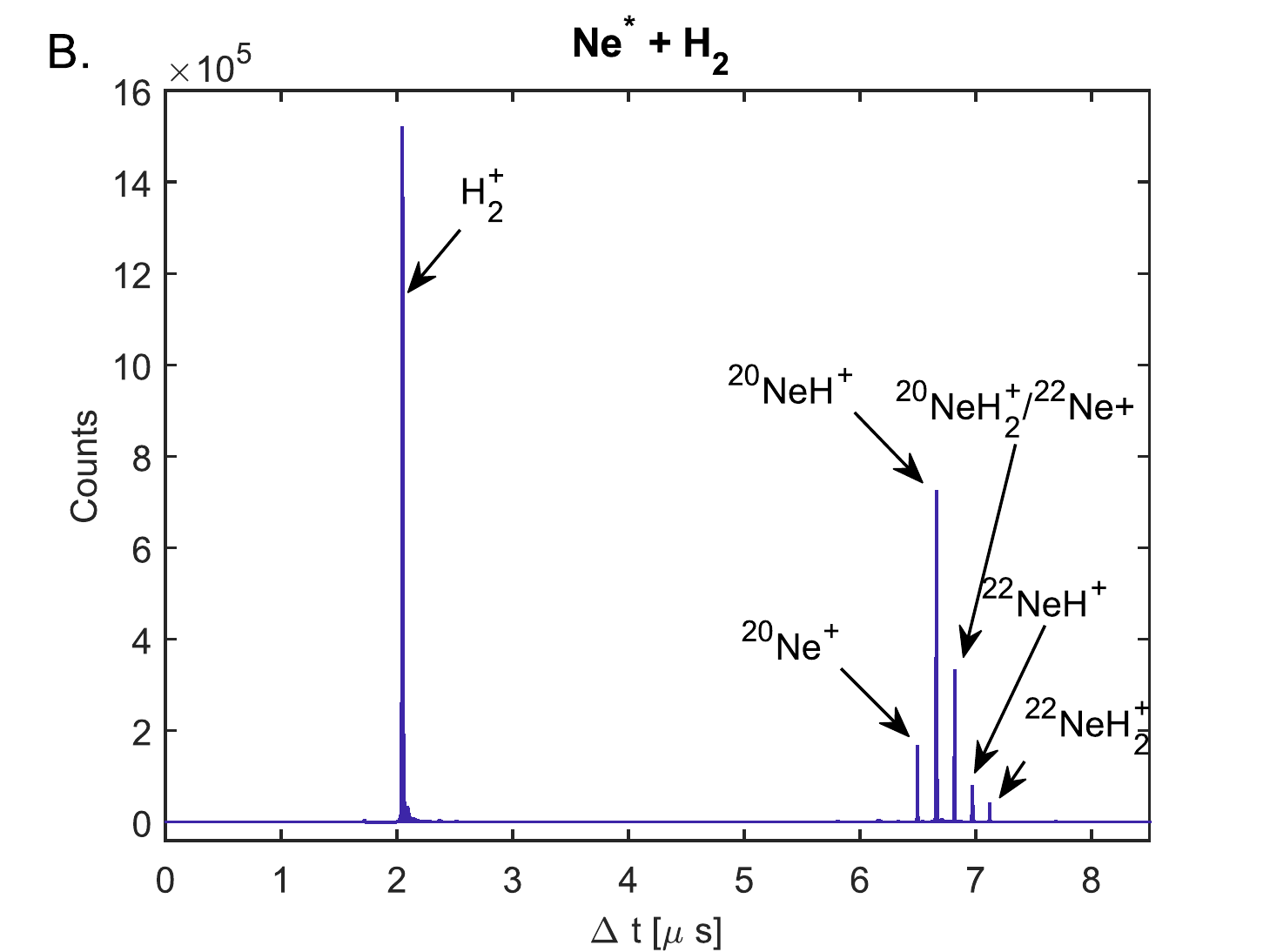}
    \caption{Mass spectrum of ionization products.}
    \label{MS_plots}
\end{figure}

The mass spectrum of ionic products was obtained by plotting the histogram of TOF difference between ions and electrons. The result is shown at Fig~\ref{MS_plots}. The main ionic product is H$_2^+$ which results from a direct Penning Ionization reaction. In addition, we observed triatomic molecular ions which result from associative ionization \cite{siska1993molecular}, NeH$^+$ and HeH$^+$ which result from proton exchange reactions between H$_2^+$ and Ne/He and Ne$^+$ and He$^+$ ions which originated from field ionization of Rydberg state atoms.\\

\noindent \underline{Slicing of VMI Images}\\

\begin{figure}[tb!]
\centering
\includegraphics[width=0.9\textwidth]{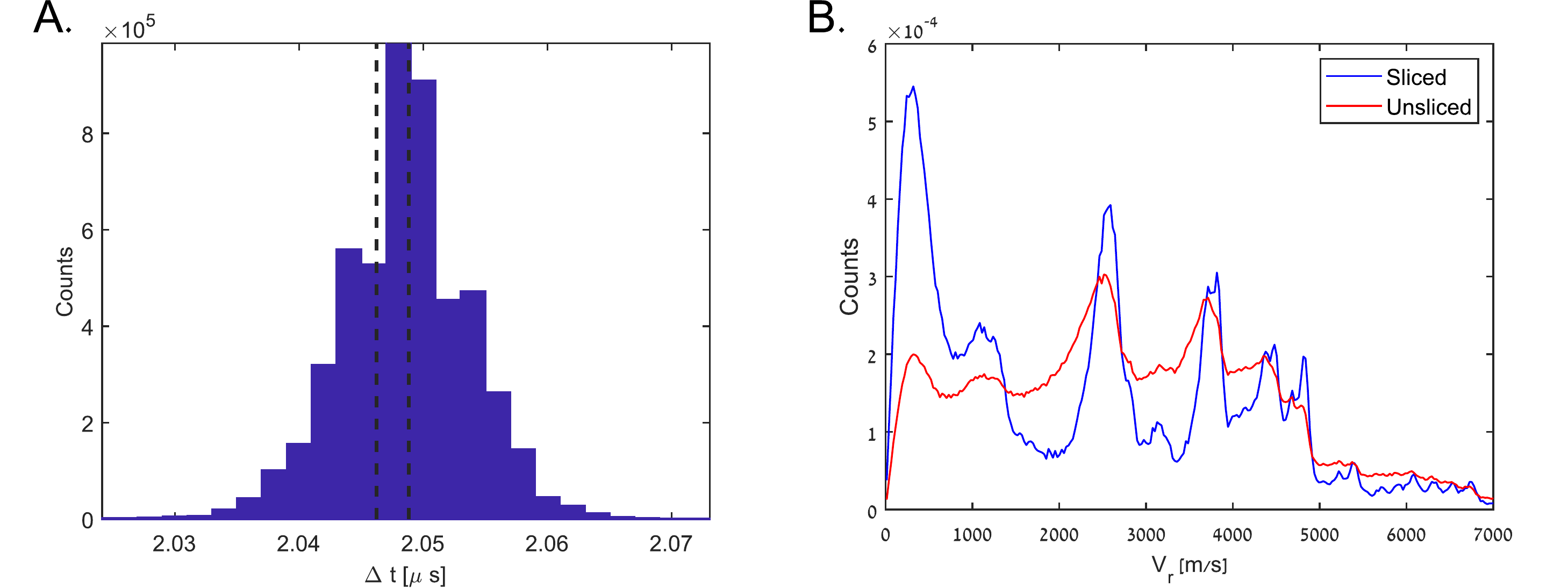}
\caption{Slicing of the H$_2^+$ ion VMI image. A. TOF distribution for H$_2^+$ ions. limits of the time-slicing are marked by a dashed black line. B. comparison of radial velocity distribution as obtained by angle integration of the sliced and unsliced VMI images.}
\label{vmi_slice}
\end{figure}

\begin{figure}[tb!]
\centering
\includegraphics[width=0.7\textwidth]{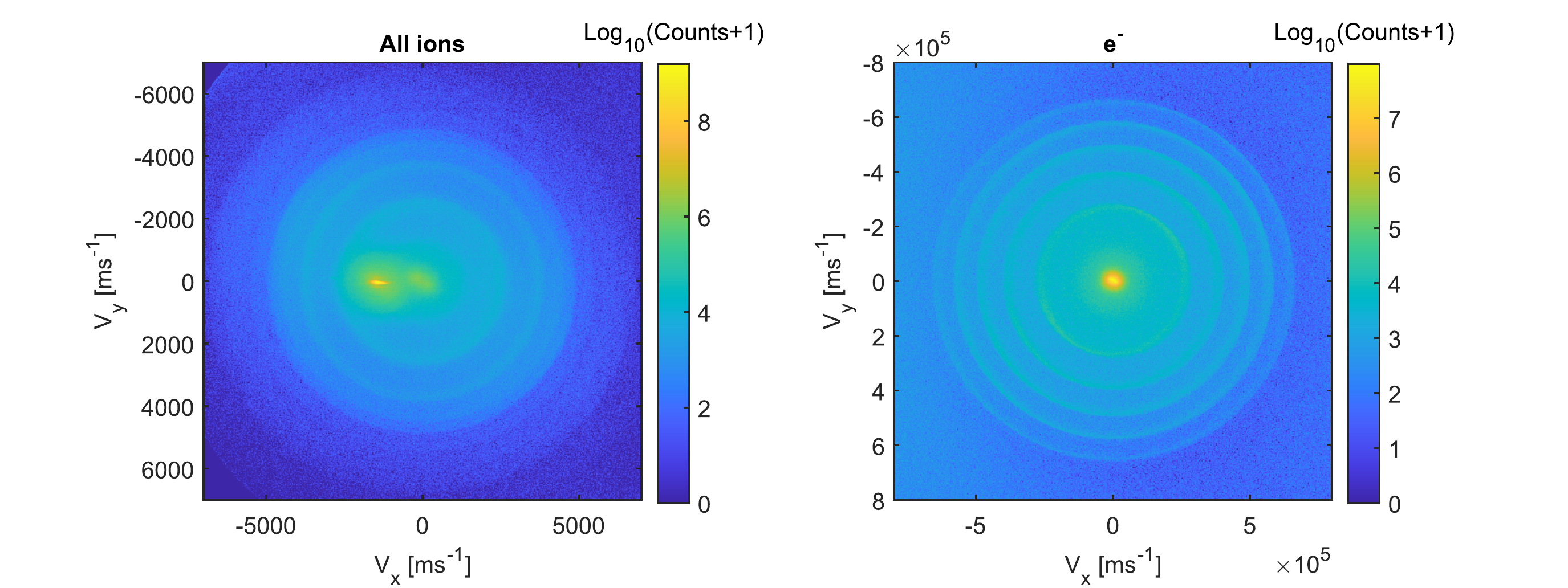} 
\includegraphics[width=0.7\textwidth]{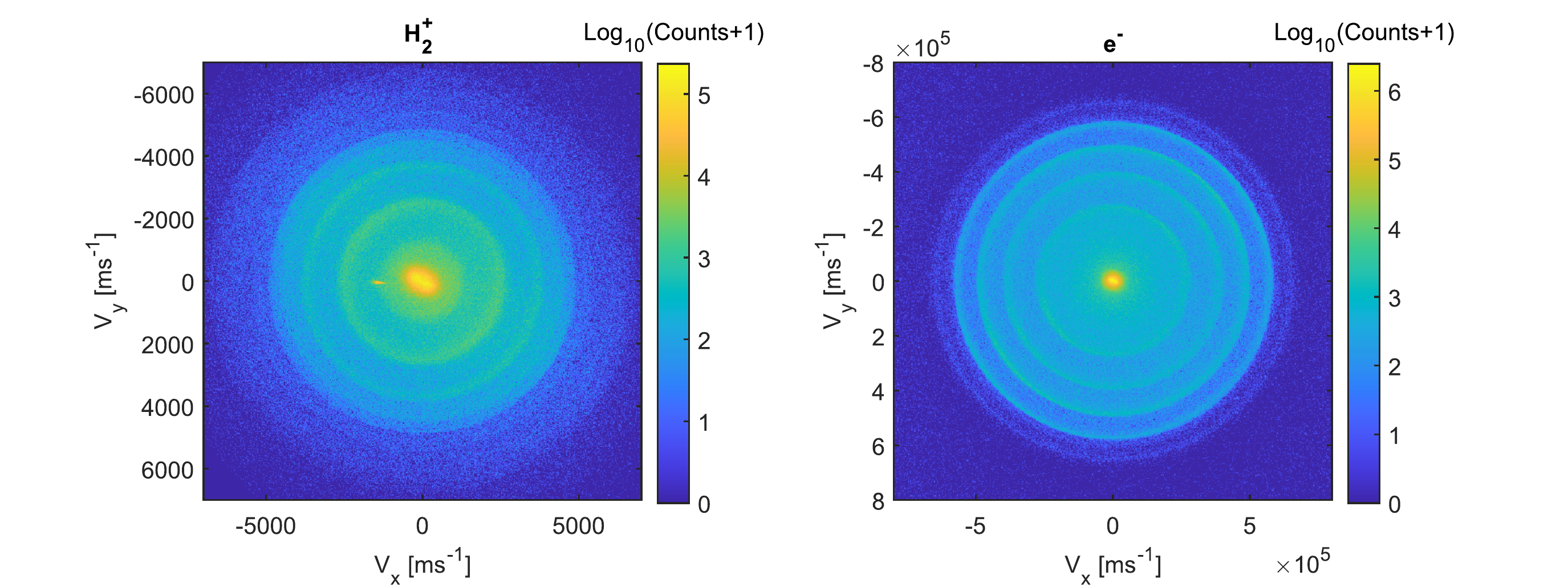} 
\includegraphics[width=0.7\textwidth]{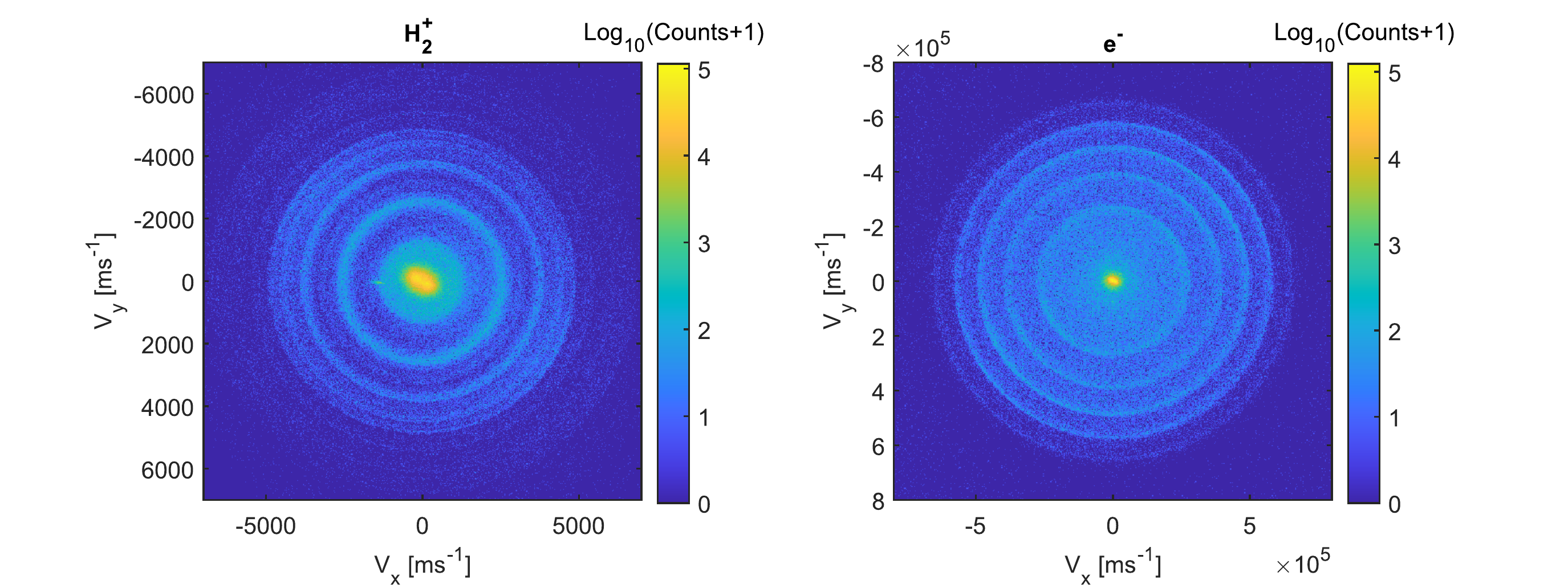} 
\caption{VMI images of ions and electrons, Ne$^*$-H$_2$ PI collisions. First row: all ions. Second row: H$_2^+$ ions and coincidence electrons. Third row: sliced images.} 
 \label{NeH2_VMI}
\end{figure}

\begin{figure}[tb!]
\centering
\includegraphics[width=0.7\textwidth]{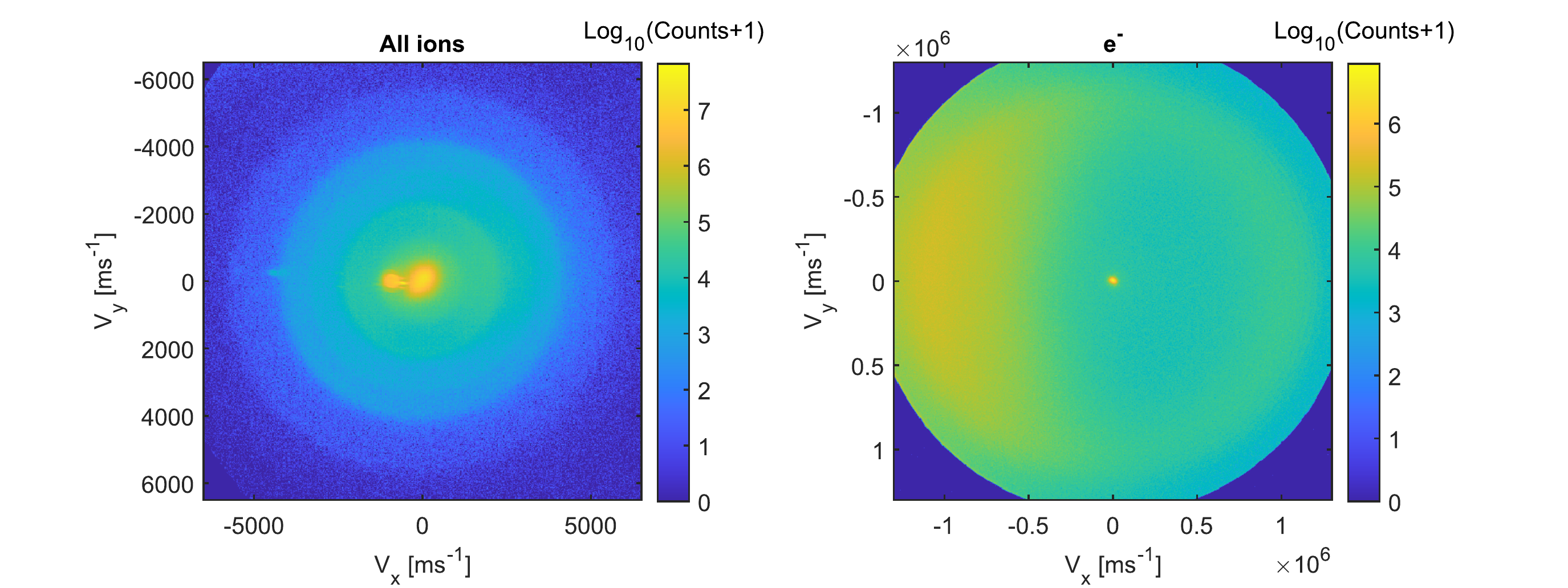} 
\includegraphics[width=0.7\textwidth]{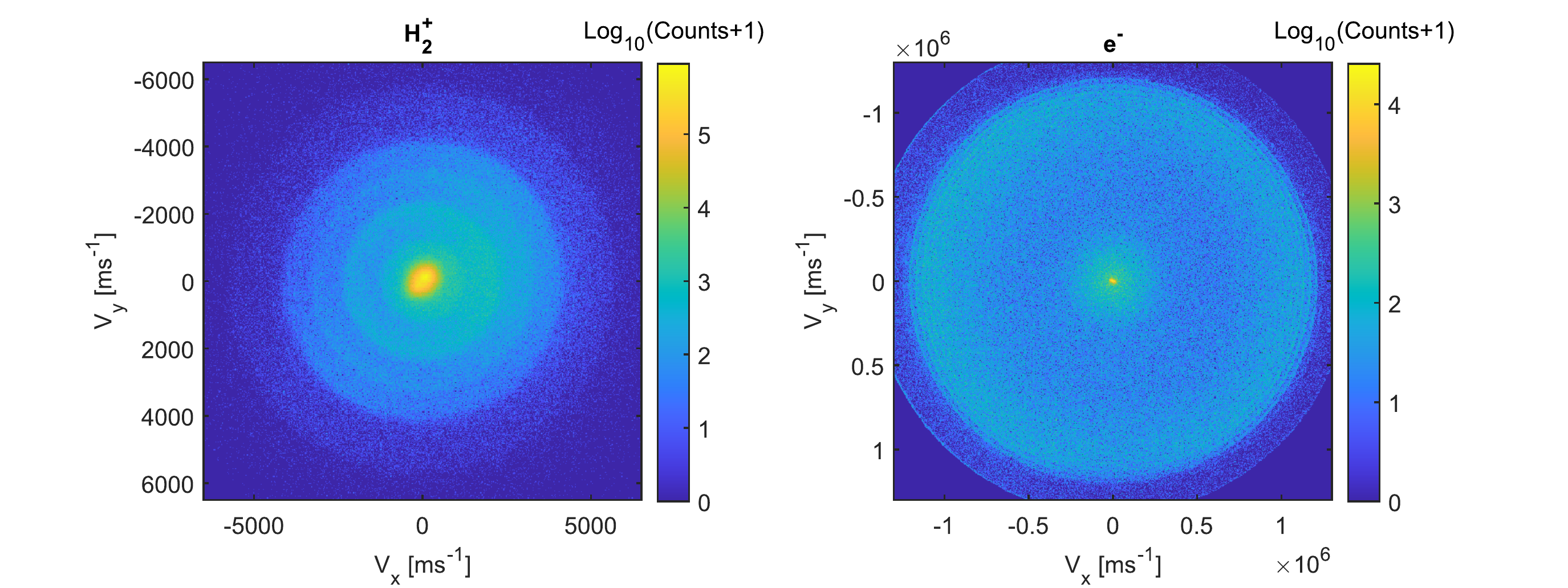} 
\includegraphics[width=0.7\textwidth]{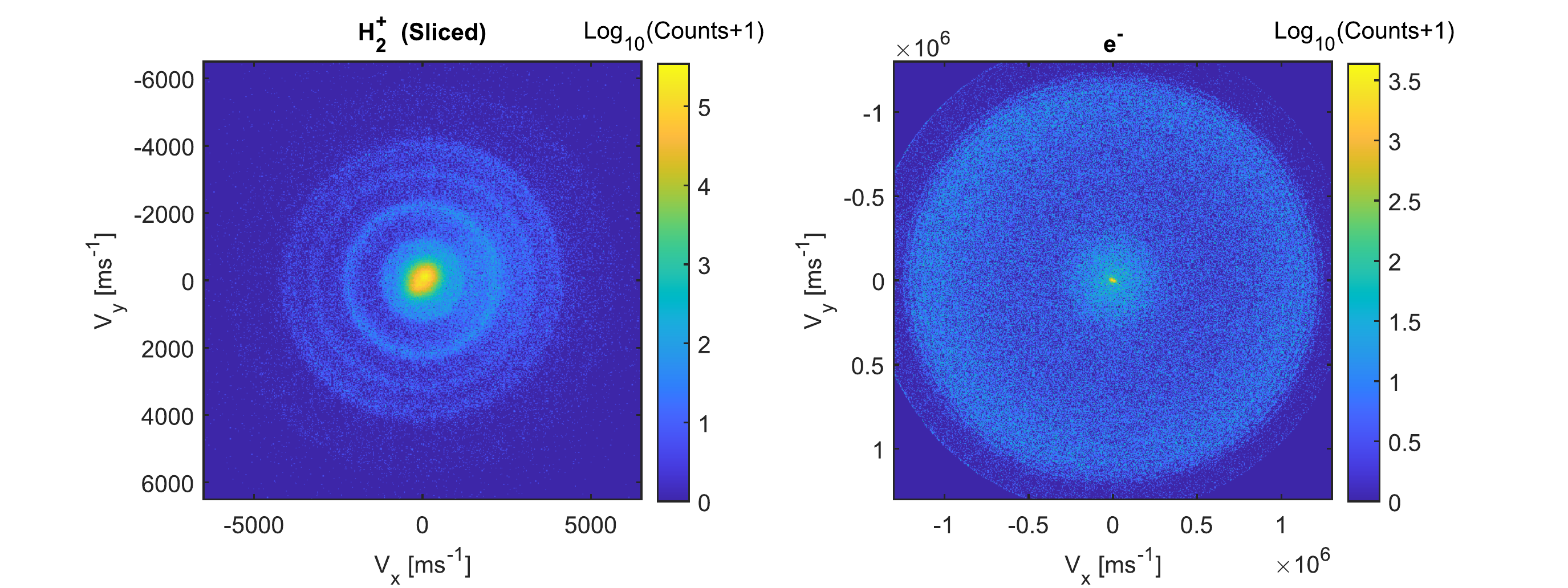} 
\caption{VMI images of ions and electrons, He$^*$-H$_2$ PI collisions.  First row: all ions. Second row: H$_2^+$ ions and coincidence electrons. Third row: sliced images.} 
 \label{HeH2_VMI}
\end{figure}

Using TOF information we were able to plot VMI images of ions of a given mass and coincidence electrons. However, the VMI technique provides only two components of the 3D velocity distribution, therefore, obtaining the distribution of velocity magnitude requires additional analysis\cite{ashfold2006imaging}. One method is based on numerically inverting the VMI image using the Abel transform, another option is to `slice' the 3D velocity distribution in the detection plane. The latter is performed by the selection of particles with zero velocity component on TOF axis. This effectively `slices' the 3D distribution around zero velocity in the direction perpendicular to the detection plane. For photoions, such a method was successfully used where zero TOF was defined by the laser pulse \cite{townsend2003direct}. Here we took a similar approach, accounting the electron arrival time as the zero time stamp. The slice width is limited by the uncertainty in the electron TOF, given as twice the turn around time in the extraction region which is in the order of 2ns for 4eV electrons. The slicing procedure for H$_2^+$ ions is shown graphically at Fig \ref{vmi_slice}. Sliced and unsliced VMI images of ions and the corresponding electronic VMI images are shown in Figs \ref{NeH2_VMI} and \ref{HeH2_VMI}.

\clearpage

\noindent \underline{Peeling of the Electronic VMI image}\\

\begin{figure}[tb!]
\centering
\includegraphics[width=0.9\textwidth]{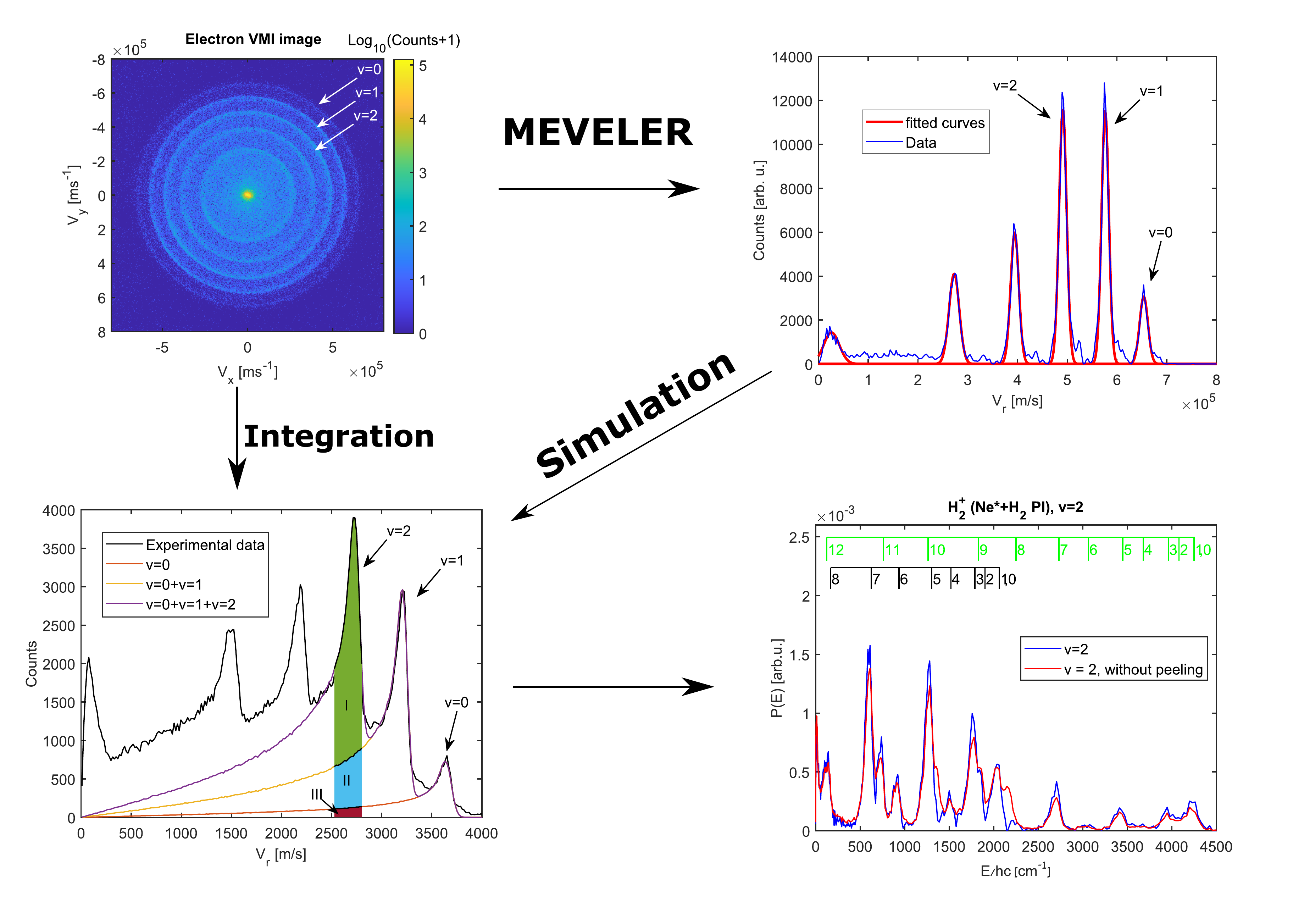}
\caption{Schematic overview of the peeling algorithm for ionic VMI image corresponding to $v=2$ initial state. Starting from the top left corner, the electron VMI image is inverted using MEVELER program to provide the radial velocity distribution (top right corner). From the latter, the contribution of individual electron energy to the 2D projected velocity distribution is simulated (bottom left corner) and presented together with radial distribution as acquired directly from the VMI image. The relative contribution of ions related to initial $v=0$ and $v=1$ is determined by the ratio of the shaded area II and III relative to the shaded area I. The ratios are used to correct the ionic image corresponding to v=2 by subtracting the contribution of VMI images corresponding to $v=1$ and $v=0$. The uncorrected and corrected ionic energy distribution for H$_2^+$ ions corresponding to the $v=2$ initial state are presented at the bottom right corner of the figure.}
 \label{peeling}
\end{figure}

\begin{figure}[ht!]
    \centering
    \includegraphics[width=1\textwidth]{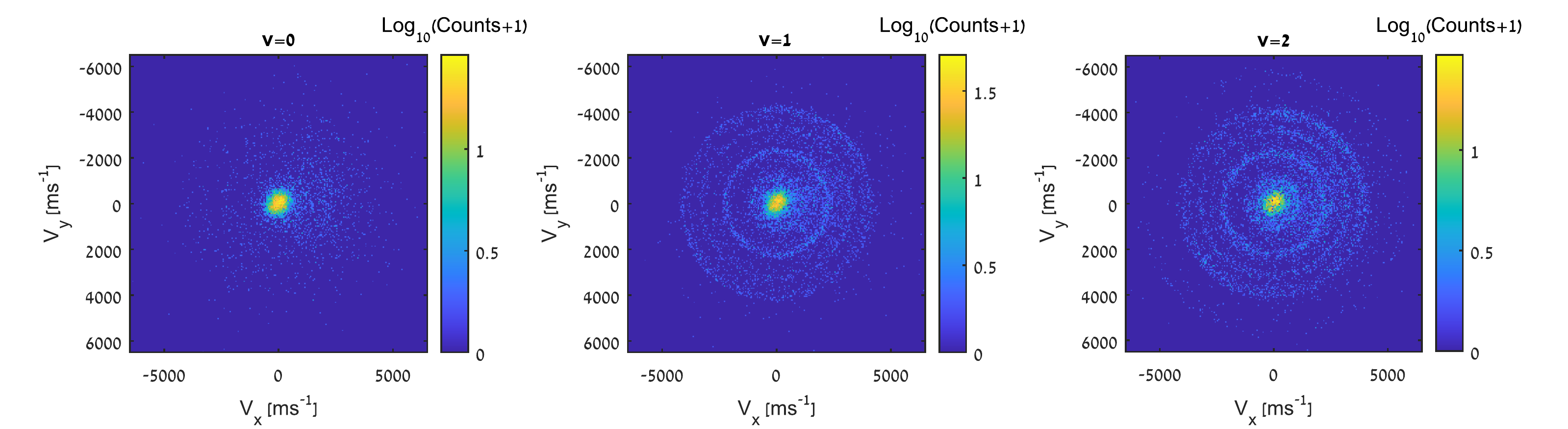} 
    \includegraphics[width=1\textwidth]{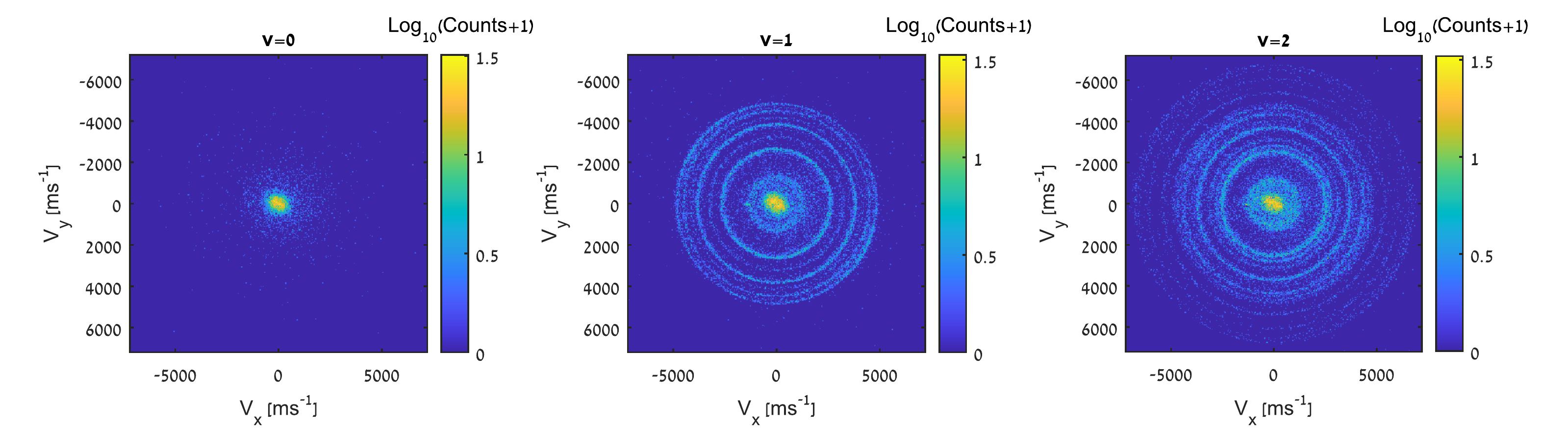}
    \caption{Sliced VMI images of H$_2^+$ per initial vibrational state. First row: He$^*$-H$_2$. Second row: Ne$^*$-H$_2$ PI collisions.}
    \label{sl_VMI_HeH2_per_v}
\end{figure}

In contrast to the ionic VMI images, the lack of a zero time stamp for the electron TOF prohibits the ability to slice the electronic VMI images. A problem arises when one wishes to perform state-to-state analysis which requires the selection of electrons in a given energy range. The presented deficiency was overcome in two steps, firstly by accounting for electrons with a minimal velocity component on TOF axis and secondly by accounting for the contribution of outer electronic energies on the inner, lower energy section of the electronic VMI image. For this purpose, we describe here a "Peeling" algorithm that corrects every ionic VMI image related to a specific electron energy due to the overlap with higher electron energies. The "Peeling" procedure is presented schematically in Fig \ref{peeling}. First, the electron VMI image is inverted using MEVELER\cite{dick2014inverting} to obtain the velocity magnitude distribution. The individual peaks are fitted to Gaussian distributions and the acquired peak position, height and width are used to simulate the projection of the individual velocity magnitude distributions back to a 2D plane. This allows us to set apart the individual contributions of each electron energy to the total electronic VMI image. For state-to-state analysis, a specific electronic radial range is chosen around the peak maximum and the coincidence ionic VMI image is obtained. Starting from the highest energy of electrons, and going inwards, the ionic VMI images from higher electron energies are subtracted from the analyzed VMI image with a ratio that is calculated according to the relative contribution of higher electronic energies.
The peeling algorithm allowed us to correctly decompose the sliced VMI images into individual images, associated to the formation of Feshbach resonances at different vibrational states of the molecular ion. The resulted VMI images are shown in Fig~\ref{sl_VMI_HeH2_per_v}.

\newpage

\noindent \underline{Theoretical methods} \\ %\label{sec:theory}

\noindent \underline{Potential Energy Surfaces}\\

\begin{figure}[tb!]
    \centering
     \includegraphics[width=0.38\textwidth]{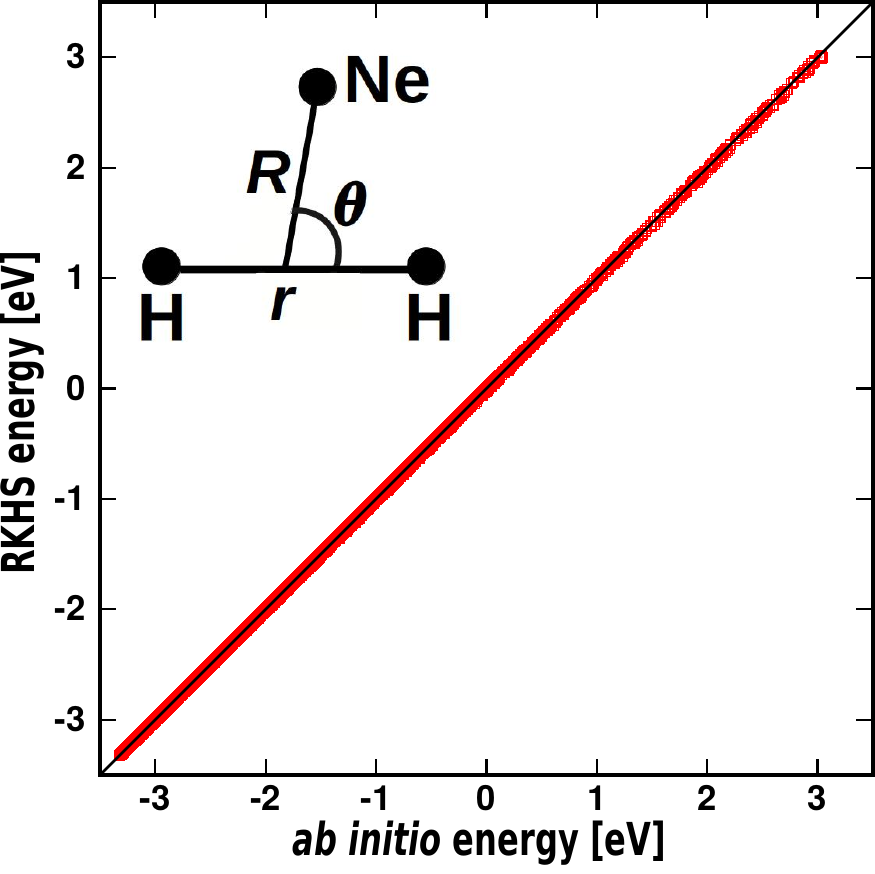}
    \caption{Correlation plot for on-grid 38200 points between the reference
  CCSD(T)/aug-cc-pV5Z \textit{ab initio} energies and
  those obtained from the RKHS with ${\rm RMSE}= 0.0014$ eV (11.3 cm$^{-1}$). The agreement between electronic structure calculations and the RKHS representation for 250 off-grid points has an RMSE of 11.8  cm$^{-1}$.}
    \label{sifig:pes}
\end{figure}

Both PESs were reproducing kernel Hilbert 
space\cite{unke2017toolkit} (RKHS) representations of reference energies computed using 
MOLPRO\cite{werner2020molpro}. For He--H$_2^+$, a full configuration interaction (FCI) PES 
generated from 7936 reference points was already available and used to investigate\cite{koner2019heh2+pes}  
near-dissociative states for which experiments had been carried out\cite{gammie2002microwave, carrington1996observation}. For Ne--H$_2^+$, a new $^2$A' PES was determined at the CCSD(T)/aug-cc-pV5Z level of theory. In this case, 38200 reference energies were  represented as a RKHS. For the Ne--H$_2^+$ PES reference energies at the CCSD(T)/aug-cc-pV5Z level of theory were determined on a grid of Jacobi
coordinates ($R$,$r$,$\theta$), where $r$ is the bond length of the
H$_2^+$ diatomic, $R$ is the distance between the center of mass
of the diatomic molecule and the rare gas atom (Ne) and $\theta$ is the angle between
$\vec{r}$ and $\vec{R}$. The grid (on-grid points) included 39 points for $r \in [1.1, 8.0]$ a$\rm _0$, 49 points for $R \in [1.0, 
45.0]$ a$\rm _0$ and 10 Gauss-Legendre quadrature points for $\theta \in [0, 90^{\circ}]$. The 
quality of the final PES is shown in Figure \ref{sifig:pes}. The root mean squared difference for energies of 250 geometries that were not used for constructing the RKHS representation (off-grid points) between reference calculations at the CCSD(T)/aug-cc-pV5Z level of theory and the evaluated RKHS representation is 11.8 cm$^{-1}$, compared with ${\rm RMSD} = 11.3$ cm$^{-1}$ for on-grid points.\\ 

\noindent \underline{Scattering calculations}\\

Our calculations model rovibrational quenching in the half-collision of the rare gas atom, He or Ne, with dihydrogen molecular ions. Since the reactive channel is not energetically accessible for vibrational excitations below $v=4$, 
the three-body Hamiltonian can be constructed using Jacobi coordinates $(R,r,\theta)$,
cf.\ Fig.~\ref{sifig:pes}.  
In the case of neon, the calculations were performed for pure $^{20}$Ne (i.e., with an atomic mass of $19.99$), which accounts for approximately $90\%$ of naturally occurring neon gas, as it was utilized in the experiment. We emphasize that our approach to the quantum dynamics is exact, i.e., we do not invoke any additional approximations beyond those made to derive the potential energy surfaces and taking the reactive channel to be inaccessible.

The total simulation consisted of two distinct steps. Firstly, the Penning ionization is modelled by diagonalizing the Hamiltonian
describing the collision of the metastable rare gas atom with neutral dihydrogen, using the adiabatic potentials~\cite{pawlak2015adiabatic,pawlak2017adiabatic,klein2017directly} for all relevant combinations of partial wave $\ell$ and total angular momentum $J$. The mapped spatial grid for the atom to diatom-COM separation consisted of 8192 points between $R=2$ and $\unit[20\,000]{a_{0}}$. 
After diagonalization, the eigenenergies that most closely correspond to the
measured experimental collision energy for each $J$, $\ell$ and spin isomer were selected.
Projecting the eigenstate for each selected eigenenergy onto the ionic surface with the double-exponential ionization probability $\Gamma(R)$ yielded the set of input wave functions for simulating the collision on the ionic surface.
The projection results in initial wave packets with spatial distributions along the atom to diatom-COM axis. The wave packets were found to be robust with regard to the energetic selection mentioned above and well approximated by a Gaussian distribution centered around $R\approx\unit[8]{a_{0}}$ for He-H$_{2}^{+}$ and $R\approx \unit[10]{a_{0}}$ for Ne-H$_{2}^{+}$, respectively.

Secondly, the half-collisions on the ionic surface
are governed by the three-body Hamiltonian,
\begin{SEquation}\label{eq:Htot}
    H_{\mathrm{tot}} = - \frac{\hbar^2}{2\mu_{\mathrm{cmplx}}} \nabla^2_{\vec R} + H_{\mathrm{ion}} + V(R,r,\theta) 
\end{SEquation}
with
\begin{SEquation}\label{eq:Hion}
    H_{\mathrm{ion}} = - \frac{\hbar^2}{2\mu_{\mathrm{ion}} } \nabla^{2}_{\vec r} %\frac{\partial^2}{\partial r^2}r + \frac{j'(j'+1)}{2 \mu_{\mathrm{ion}} r^2} 
    + V_{\mathrm{ion}}(r)\,,
\end{SEquation}
where $\mu_{\mathrm{cmplx}}$ and $\mu_{\mathrm{ion}}$ are the reduced masses of the Rg-H$_2^+$ complex and the H$_2^+$ molecule ion, respectively. The Hamiltonian~\eqref{eq:Htot} conserves parity as well as the total angular momentum  $\vec J=\vec j+\vec L$, where $\vec j$ the describes molecular rotation and $\vec L$ is the angular momentum of the rare gas atom relative to the molecule. It is thus convenient to represent the eigenfunctions of $H_\mathrm{{tot}}$ in the coupled angular basis \cite{pack1973space},
\begin{SEqnarray}\label{eq:Psitot}
    \Psi^{JMj\ell v}(\vec{R},\vec{r}) &=& \frac{1}{R} \sum_{j',\ell',v'} G^{Jj\ell v}_{j'\ell'v'}(R) \frac{\chi_{j'v'}(r)}{r}\\
    && \mbox{} \times \sum_{m_{j}=-j}^{j} \sum_{m_{\ell}=-\ell}^{\ell} C^{JM}_{m_{j}m_{\ell}}\, Y_{jm_{j}}\left(\theta_r,\varphi_r\right)\,Y_{\ell m_{\ell}}\left(\theta_R,\varphi_R\right),
    \nonumber
\end{SEqnarray}
where $C^{JM}_{m_{j}m_{\ell}}$ are the Clebsch-Gordan coefficients for the transformation from the eigenfunctions of the uncoupled angular momenta $\hat{j}^2,\hat{j}_z,\hat{L}^2,\hat{L}_z$ (with quantum numbers $j,\ell,m_j,m_{\ell}$) to the eigenfunctions of $\hat{j}^2, \hat{L}^2, \hat{J}^2, \hat{J}_z$ (with quantum numbers $j,\ell,J,M$). The rovibrational eigenstates of the molecular ion, $\chi_{jv}(r)$, with vibrational quantum number $v$ were obtained by diagonalizing $H_{\mathrm{ion}}$ using a discrete variable representation. Furthermore, in Eq.~\eqref{eq:Psitot}, 
$G_{j'\ell'v'}^{Jj\ell v}(R)$ is the radial wave function in the ($j',\ell',v'$)-exit channel obtained from an entrance channel $j,\ell,v$. In order to represent the interaction potential (given in Jacobi coordinates) in the coupled angular basis, we employed an expansion into Legendre-polynomials in $\theta$ and $r$, 
\begin{SEquation}
    V(R,r,\theta) = \sum_{k=0}^{\infty}\sum_{n=0}^{\infty} V_{n,k}(R) P_{k}(\cos\theta) \tilde{P}_{n}(r)\,,
\end{SEquation}
where $\tilde{P}_{n}(r)$ is the $n^{\mathrm{th}}$ order Legendre polynomial including the rescaling of $r$ to $[-1,1]$.
This expansion results in $R$-dependent matrix elements of $V$ in the channel basis,
\begin{SEquation}\label{eq:V}
    \bra{Jj'\ell'v'}V\ket{Jj''\ell''v''} = \sum_{k=0}^{\infty} \sum_{n=0}^{\infty} V_{n,k}(R) 
    \bra{Jj'\ell'}P_{k}(\cos\theta)\ket{Jj''\ell''}
    \int_{\mathbbm{R}_{+}}\chi_{j'v'}^{*}(r) \tilde{P}_{n}(r) \chi_{j''v''}(r)\,dr
\end{SEquation} 
and the eigenvalue problem of $H_{\mathrm{tot}}$ takes the form
\begin{SEquation}\label{eq:CC}
    \left[\frac{\partial^2}{\partial{R^2}} + k^{2}_{j'v'} - \frac{\ell'(\ell'+1)}{R^2}\right] G_{j'\ell'v'}^{Jj\ell v}(R) = \frac{2\mu_{\mathrm{cmplx}}}{\hbar^2} \sum_{j''\ell''v''} \bra{Jj'\ell'v'}V\ket{Jj''\ell''v''} G_{j''\ell''v''}^{Jj\ell v}(R)\,,
\end{SEquation}
where $k_{j'v'}^{2} = \frac{2\mu_{\mathrm{cmplx}}}{\hbar^2} (E-\mathcal{E}_{j'v'})$ with $E$ the total energy and $\mathcal{E}_{j'v'}$ the dihydrogen rovibrational energy levels. 

We solved the coupled channels equations~\eqref{eq:CC} using the renormalized Numerov method~\cite{johnson1978renormalized} for the ratio square or ${\bm Q}$-matrices which relate the wave function on neighboring points of the radial grid~\cite{gadea1997nonradiative}, propagating them outward from $R_{\min} = \unit[1.0]{a_0}$ to $R_{\max} = \unit[100]{a_0}$. Outgoing wave boundary
conditions are imposed to find the wave function at $R=\unit[100]{a_0}$. Stored
${\bm Q}$-matrices could then be used to propagate inward to compute the scattering wave functions on the grid~\cite{gadea1997nonradiative}. The wave packet $\phi$ describing the Penning ionization obtained in the first step as explained above would then be accounted for by projecting onto a set of eigenstates of $H_{\mathrm{tot}}$. 
However, in order to avoid storing ${\bm Q}$-matrices, we instead propagate the wave packet outward together with the Q-matrices~\cite{gadea1997nonradiative}. 
For each energy $E$, the ${\bm K}$-matrix can be obtained from the flux-normalized boundary conditions. Once the propagated input wave packet $\phi_{j''\ell'' v''}^{Jj\ell v}(R)$ has been obtained at $R=R_{\mathrm{max}}$, the integral cross section for each exit channel $j'\ell' v'$ can be evaluated as
\begin{SEquation}\label{eq:CS}
    \sigma_{j'\ell'v'}^{Jj\ell v} = \frac{1}{2\pi} \sum_{j''\ell''v''} \Big| A_{j'\ell'v';j''\ell'' v''} \phi_{j''\ell''v''}^{Jj\ell v}(R_{\max})\Big|^2\,\,,
\end{SEquation}
where 
\begin{SEquation}
    {\bm A} = 2i[{\bm F}(R_\mathrm{max})-{\bm G}(R_\mathrm{max}){\bm K}]({\bm 1}+i{\bm K})^{-1}    
\end{SEquation}
 with ${\bm F}$ and
${\bm G}$ being diagonal matrices of the asymptotic boundary conditions, given by spherical Bessel functions of the first and second kind, respectively, and ${\bm K}$ is the hermitised ${\bm K}$-matrix.

We have used 4002 energies $E$  ranging from $\unit[200]{cm^{-1}}$ below to $\unit[200]{cm^{-1}}$ above the 
dissociation threshold of the given entrance channel. 
Such a fine energy sampling ensured that
Feshbach resonances, which can have narrow spectral widths, were adequately captured.
For each energy, cross sections were obtained separately for each channel,
%characterized by quantum numbers $v$, $j$, $\ell$ 
with the difference in the collision energy and the channel's internal energy yielding the fragment kinetic energy. Evaluating the total cross section over all outgoing channels, as shown in Figs.~2C and~4 of the main text, all individual contributions to the collision cross section were added incoherently both in terms of channels and energy, cf. Eq.~\eqref{eq:CS}. This implies in particular that no coherence between different Feshbach resonant states is assumed, which is in line with the experimental observations. The finite-basis representation included 31 terms for both the diatomic separation $r$ and the polar angle $\theta$.
Consequently, the simulation was performed in a coupled channel basis consisting of the vibrational quantum numbers $v$ and the uncoupled angular momenta $j$ and $\ell$.
The graphs in Fig.~4 were obtained by creating histograms in kinetic energy for all channels and total angular momenta $J$ ranging from 0 to 6 with corresponding entrance channel partial waves $\ell$ multiplied by their respective weights. Our calculations naturally include the elastic pathway where the initial vibrationally excited state remains unchanged. The contribution is observed as an incline towards zero energy, starting around 40 cm$^{-1}$. The same contribution to the scattering flux is shown in Fig.~2A as the positive energy part of the overlap plots.
As seen in the overlap plots in Fig.~2, the resonance width of individual Feshbach resonances ranges between 1cm$^{-1}$ to 10cm$^{-1}$. The overall natural peak width which includes a few resonance states from a few angular momentum states results in a width of around 20cm$^{-1}$.
For comparison with the experiment, the cross sections as functions of the kinetic energy have been convoluted taking the resolution of the detector into account.
To this end, the experimental resolution per particle energy is directly simulated, taking into account the spread in COM velocity, the VMI resolution, and the limit of the time-slicing procedure. Overall, the resolution ranges between 10 cm$^{-1}$ for the lowest energy peaks to 30 cm$^{-1}$ for the highest energy peaks in Fig. 4A. %between 25 and 50 m/s.
Scattering wave functions as shown in Fig.~2 of the main text were constructed from the Numerov ${\bm Q}$-matrices.
% by iterating backwards from the asymptotic solution at the end of the spatial grid. 

\newpage
\subsection*{Supplementary text}

\noindent \underline{Comparison to EPDS} \\

It is interesting to compare our method to electron photodetachment spectroscopy (EPDS)~\cite{kim2015spectroscopic,otto2014imaging}. The most significant difference is that we obtain correlated energy distributions of all charged products and not just of the electrons. This allows us to obtain both the resonance energy and the final state distribution. However, contrary to ionization by a laser, we have no experimental time stamp to when ionization occurs. As a result, we cannot slice the electron VMI images. To overcome this deficiency, we have developed a 'peeling' algorithm which is presented above. Ionization by collisions has an additional implication on the imaging system design due to the large ionization volume. This results in aberrations from particles moving close to the VMI orifice. 
Another difference is that there is no tunability to the level of excitation. The initial excitation energies are given by the energy of the metastable states of the noble gas atom. For collisions with He*, this results in electrons with up to 4$\,$eV kinetic energy. In EPDS and specifically in SEVI (slow photoelectron velocity map imaging), the excitation is tuned so that the ejected electron has lower energy and is therefore easier to detect with higher resolution. However, since our experiment is based on cold collisions, the electron resolution in principle can be further increased, possibly enabling the detection of individual Feshbach resonance states.\\

\noindent \underline{Comparing theory to experimental data} \\

\begin{figure}[t]
    \centering
    \includegraphics[width=0.48\textwidth]{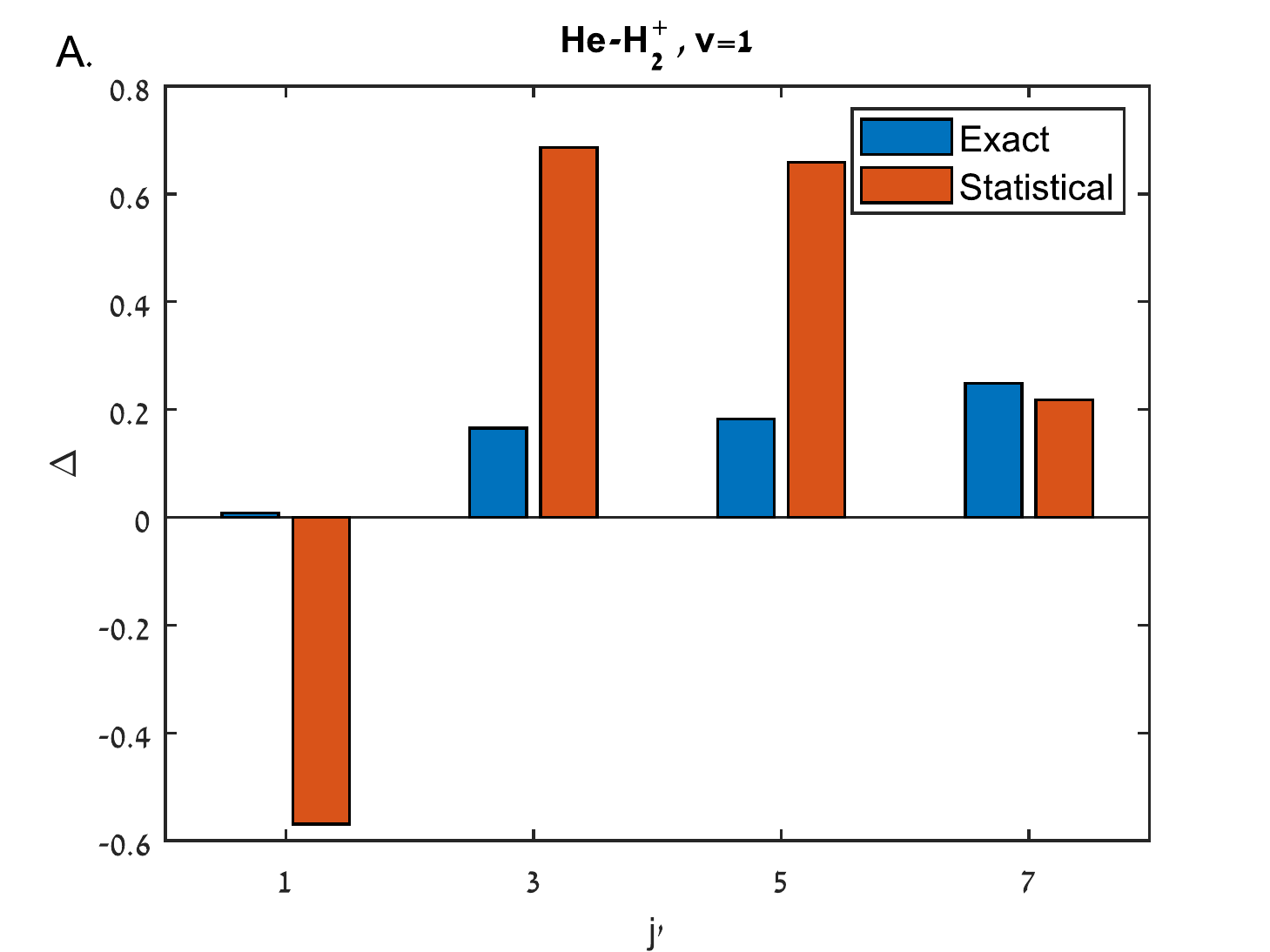} 
    \includegraphics[width=0.48\textwidth]{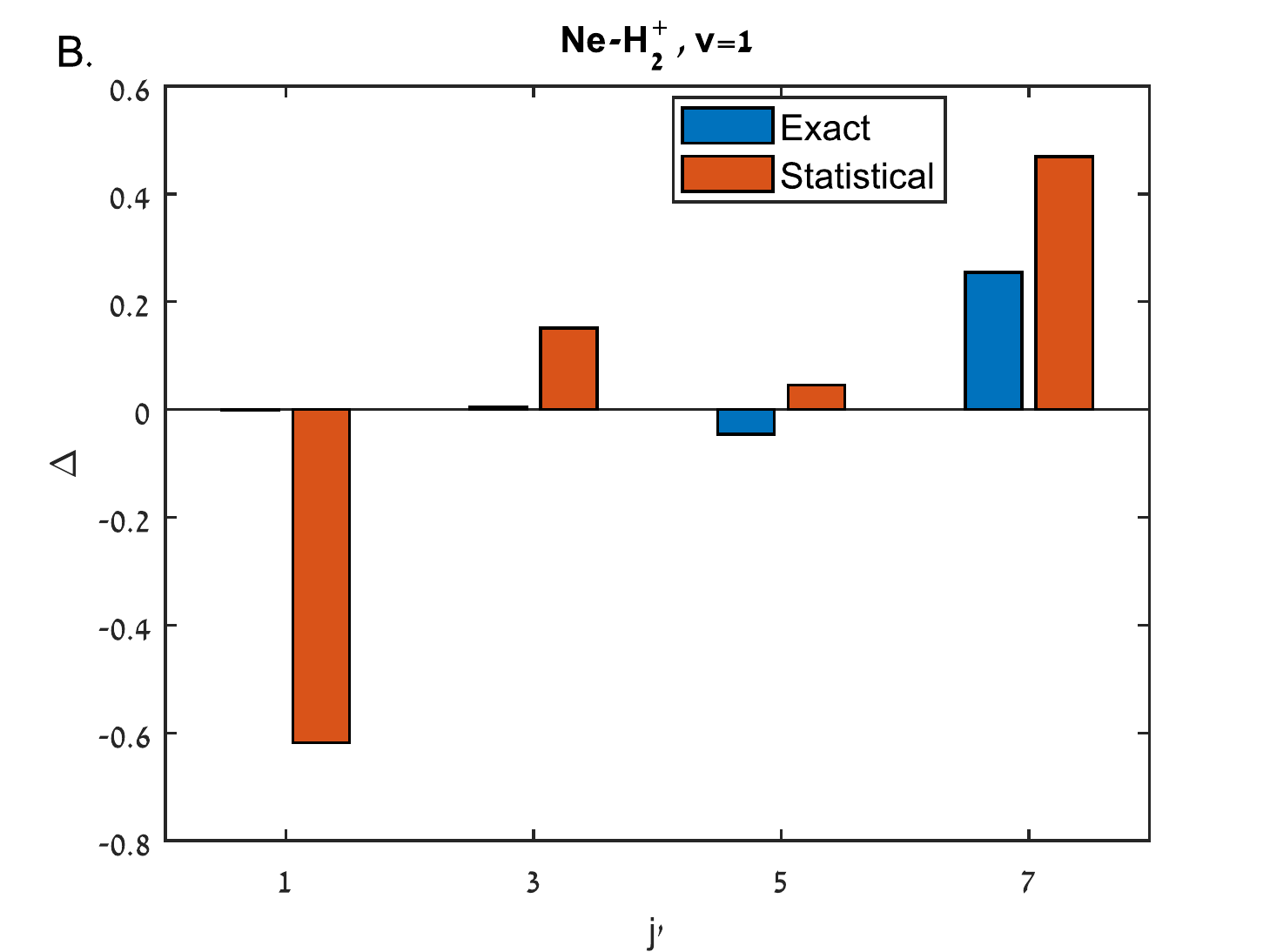}
    \caption{Quantitative comparison of energy distributions obtained by the statistical model and exact calculations to the experimental data for initial $v=1$ and ortho di-hydrogen. The height of individual bars represents the difference in peak area, compared to the experimental peak area. }
    \label{comp_st_exp}
\end{figure}

We present in Fig.~\ref{comp_st_exp} a comparison between the experimental data and theoretical results obtained in a two-fold way, once by exact coupled channels calculations on an \textit{ab initio} potential energy surface, as described above, and second using a statistical model as described in the main text. 
We quantify the contrast between theoretical and experimental results by calculating the difference in peak area,
\begin{SEquation}
    \Delta=\frac{\int_a^b P_{th}(E)dE-\int_a^b P_{exp}(E)dE}{\int_a^b P_{exp}(E)dE}
\end{SEquation}
where $P_{th}(E)$ and $P_{exp}(E)$ are the theoretical and experimental energy distributions. Integration limits $a$ and $b$ were defined as plus/minus two standard deviations from the peak center, as obtained by fitting the experimental data to a Gaussian function. We chose to analyze the data for ortho di-hydrogen where the signal to noise ratio is the highest.
In Fig.~\ref{comp_st_exp}, the overall agreement of the exact quantum mechanical calculations with the experimental data is evidently much better than for the statistical model. Moreover, for the statistical model, we observe a trend in the disagreement of underestimating decay to states with low final $j'$ and overestimating decay to states with high final $j'$. 

%\newpage
%\bibliographystyle{ieeetr}
%\bibliography{suprefs}

%\end{document}

\end{document}